\def\frac#1#2{{{{#1}}\over{{#2}}}}

\documentclass[12pt]{article}
\usepackage{wasysym, xcolor}

\newsavebox{\ns}
\newsavebox{\dbrane}
\newsavebox{\dbshort}

\usepackage{epsfig}
\usepackage{amssymb}

\def\appendix{{\newpage\section*{Appendix}}\let\appendix\section%
        {\setcounter{section}{0}
        \gdef\thesection{\Alph{section}}}\section}

\newcommand\ba{\begin{eqnarray}}
\newcommand\ea{\end{eqnarray}}

\def\ccr{c}
\def\ccl{\tilde{c}}
\newcommand{\nn}{\nonumber}

\def\Dslash{\,\,{\raise.15ex\hbox{/}\mkern-12mu D}}
\def\Dbarslash{\,\,{\raise.15ex\hbox{/}\mkern-12mu {\bar D}}}
\def\delslash{\,\,{\raise.15ex\hbox{/}\mkern-9mu \partial}}
\def\delbarslash{\,\,{\raise.15ex\hbox{/}\mkern-9mu {\bar\partial}}}
\def\pslash{\,\,{\raise.15ex\hbox{/}\mkern-9mu p}}
\def\calDslash{\,\,{\raise.15ex\hbox{/}\mkern-12mu {\cal D}}}

\newcommand{\hh}{{1\over 2}}

\renewcommand{\ll}{_}
\newcommand{\uu}{^}
\newcommand{\pp}{\partial}

\renewcommand{\exp}[1]{{\rm exp}\left \{#1 \right \}}

\renewcommand{\d}{\delta}

\renewcommand{\dag}{{}^\dagger{}}

\newcommand{\s}{\sigma}
\renewcommand{\t}{\tau}

\newcommand{\sqd}{^2}

\renewcommand{\hh}{{1\over 2}}

\newcommand{\eee}[1]{\ba{#1}\ea}

\renewcommand{\t}{\tau}

\renewcommand{\b}{\beta}

\newcommand{\llsk}{\hskip .5in}

\newcommand{\st}{{}^*}
\newcommand{\D}{\Delta}

\newcommand{\pr}{^\prime {}}

\newcommand{\IZ}{\relax\ifmmode\mathchoice
{\hbox{\cmss Z\kern-.4em Z}}{\hbox{\cmss Z\kern-.4em Z}}
{\lower.9pt\hbox{\cmsss Z\kern-.4em Z}} {\lower1.2pt\hbox{\cmsss
Z\kern-.4em Z}}\else{\cmss Z\kern-.4em Z}\fi} \font\cmss=cmss10
\font\cmsss=cmss10 at 7pt
\newcommand{\inbar}{\,\vrule height1.5ex width.4pt depth0pt}
\newcommand{\IC}{{\relax\hbox{$\inbar\kern-.3em{\rm C}$}}}
\newcommand{\IQ}{{\relax\hbox{$\inbar\kern-.3em{\rm Q}$}}}
\newcommand{\IH}{{\relax\hbox{$\inbar\kern-.3em{\rm H}$}}}
\newcommand{\IP}{\relax{\rm I\kern-.18em P}}

\newcommand{\ed}{\dot{e}}

\renewcommand{\pr}{{}^\prime{}}

\newcommand{\IR}{\relax{\rm I\kern-.18em R}}
\def\blfootnote{\xdef\@thefnmark{}\@footnotetext}

\newcommand{\bm}{\begin{matrix}}
\renewcommand{\em}{\end{matrix}}

\newcommand{\lno}{\left .}
\newcommand{\rno}{\right .}

\newcommand{\rba}{\right |}

\newcommand{\upp}[1]{^{({#1})}{}}

\newcommand{\co}{{\cal O}}

\newcommand{\rr}[1]{(\ref{{#1}})}
\newcommand{\bbb}{\ba\begin{array}{c}}
\renewcommand{\eee}{\nonumber\end{array}\ea}
\newcommand{\een}[1]{\label{#1}\end{array}\ea}

\def\hilo{{}_{{}_{{}_{{}_{{}_{}}}}} {}^{{}^{{}^{}}}}

\newcommand{\heading}[1]{\begin{center}\it {#1} \rm \end{center}}

\def\lrdd{\left ( ~}
\def\rrdd{\hilo \right )}
\def\lsqq{\left [ ~}
\def\rsqq{\hilo \right ]}

\newcommand{\kket}[1]{\left | {#1} \right \rangle }

\def\bi{\begin{itemize}}
\def\ei{\end{itemize}}

\def\ed{\end{document}}

\renewcommand{\rr}[1]{(\ref{#1})}

\def\xxn{\\ \\}
\def\xxx{\nn\\ \nn\\}

\def\tb{\bar{\tau}}

\def\ctot{c_{\rm total}}
\def\uwu{^{(\rm warm-up)}}
\newcommand{\aaa}[1]{}
\def\FIXITT{\Lambda}
\newcommand{\edA}[1]{{#1}}
\newcommand{\subA}[2]{{#1}}
\definecolor{Red}{rgb}{1,0,0}
\newcommand{\nts}[1]{}

\newcommand{\lsim}{\mathrel{\hbox{\rlap{\lower.55ex
\hbox{$\sim$}} \kern-.3em \raise.4ex \hbox{$<$}}}}
\newcommand{\gsim}{\mathrel{\hbox{\rlap{\lower.55ex
\hbox{$\sim$}} \kern-.3em \raise.4ex \hbox{$>$}}}}

\begin{document}

\begin{titlepage}
\begin{flushright}
%PREPRINT-NUMBER\\
IPMU-09-0022\\
\end{flushright}
\vspace{15 mm}
\begin{center}
  {\Large \bf  A Universal Inequality \\ for \\ \vspace{2 mm} CFT and Quantum Gravity
%\\ \vspace{.11in}
%to the duality web
}
\end{center}
\vspace{6 mm}
\begin{center}
{ Simeon Hellerman }\\
\vspace{6mm}
{\it Institute for the Physics and Mathematics of the Universe\\
The University of Tokyo \\
 Kashiwa, Chiba  277-8582, Japan}
\end{center}
\vspace{6 mm}
\begin{center}
{\large Abstract}
\end{center}
\noindent
%%%%%%%%%%%%%%%%%%%%%%%%%%%%%%%%%%%%%%%%%%%%%%%%%%%%%%%%%%%%%%%%%%%%%%%%%%%%%%%%%%%%
We prove that every unitary
two-dimensional conformal field theory (with no extended chiral algebra, and with $\ccr, \ccl > 1$) contains a primary operator with dimension
$\D\ll 1$ that satisfies $0 < \D\ll 1 < {{\ccr + \ccl}\over{12}} + 0.473695$.
Translated into gravitational language using the AdS$_3$/CFT$_2$ dictionary, our result proves rigorously that the lightest massive excitation in \it any \rm theory of 3D matter and gravity with cosmological constant $\FIXITT < 0$ can be no heavier than $1/(4 G_N) + o ( \sqrt{-\FIXITT} ) $.  In the flat-space approximation, this limiting mass
is twice that of the lightest BTZ black hole.  The derivation applies at finite central charge for the boundary CFT, and does not rely on an asymptotic expansion at large central charge.  Neither does our proof rely on any special property of the CFT such as supersymmetry or holomorphic factorization, nor on any bulk interpretation in terms of string theory or semiclassical gravity.  Our only assumptions are unitarity and modular invariance of the dual CFT.  Our proof demonstrates for the first time that there exists a universal center-of-mass energy beyond which a theory of "pure" quantum gravity can never consistently be extended.

%%%%%%%%%%%%%%%%%%%%%%%%%%%%%%%%%%%%%%%%%%%%%%%%%%%%%%%%%%%%%%%%%%%%%%%%%%%%%%%%%%%%%
\vspace{1cm}
\begin{flushleft}
\today
%September 16, 2007
\end{flushleft}
\end{titlepage}
\tableofcontents
\newpage

\section{Introduction}

Quantum gravity in three dimensions \cite{threedqg1, threedqg2, threedqg3, threedqgchernsimonstownsend, threedqgchernsimonswitten} has long been a subject of much interest. Particularly interesting is the case of three dimensional quantum gravity with negative cosmological constant, which has anti-de Sitter space (AdS) as a maximally symmetric solution.
Studying the case of negative cosmological constant allows us to confront specific treatments of the quantum dynamics with a set of general principles \cite{wittenadscft} which any consistent theory of quantum gravity in AdS is believed to obey.  Namely, any quantum mechanical model of gravity in AdS must have its dynamics encoded by a dual theory on the boundary of the spacetime.  Furthermore this dual theory should be a conformal field theory in one dimension less than that of the bulk spacetime, which satisfies the usual axioms of unitarity, locality, the existence of an operator product expansion, and so on.

This duality, known as the AdS/CFT correspondence, has had many applications, but in this paper we wish to exploit a particular one of its virtues, namely its role as a \it universal set of rules \rm for consistent quantum gravity.  The correspondence reduces a very complicated, badly understood and seemingly ill-defined set of theories -- namely, models of quantum gravity in D dimensions -- to the precisely defined set of (D-1)-dimensional CFT.  This allows us in principle to make definite statements about models of quantum gravity, and in particular to rule out the possibility of quantum gravity theories with certain
hypothetical properties.  For instance, unitary CFT in two dimensions with with central charge greater than 1 must contain an infinite number of conformal families.  On the quantum gravity side, this tells us that \edA{in} a consistent theory of quantum gravity, the spectrum of states cannot be
accounted for solely by excitations of the metric.  There must exist massive states in addition to the boundary graviton gas.  In the limit where the energy is high compared to the Planck mass $1/G_N$, the density of such states is predicted by Cardy's formula \cite{cardy} and agrees with the geometric prediction for the Bekenstein-Hawking entropy of the AdS/Schwarzschild black hole in the r\'egime where the approximation by semiclassical general relativity is valid \cite{stromthermal}.  These constraints are central to the study of the fundamental degrees of freedom of quantum gravity and of their dynamics.

Nonetheless, the predictions of Cardy's formula are in some sense unpalatably weak: massive states in AdS will appear -- eventually, at some energy -- and assume a particular entropy -- approximately, in an approximation that will eventually be good at high enough temperatures.  The predictions derived from Cardy's formula suffer from the problem of \it asymptoticity -- \rm they are asymptotic predictions that can never be falsified by performing experiments at a given energy scale or a given temperature.  Cardy's formula is not sufficient to falsify the \subA{consistency of a candidate spectrum}{existence of a dual CFT}, no matter how high the energy \subA{level one may examine}{scale of an experiment}: the formula gives precise information about the behavior of the level densities at sufficiently high energies, but remains completely silent as to the energy threshold at which the asymptotic predictions begin to apply.

Meanwhile, if we would like to derive a firm prediction for the lowest center-of-mass energy at which new states must appear in a particular theory, we cannot usefully apply methods of effective field theory in the bulk: Planck-scale black holes have quantum corrections to their masses that are necessarily of the same order of magnitude as the semiclassical prediction.

Recently, an intriguing paper appeared \cite{threedqgadscftwitten} (based on earlier work \cite{hohnone, hohntwo}) proposing a context in which this question can be addressed with great precision.  The paper \cite{threedqgadscftwitten} examines the gravity duals of two dimensional CFTs in which the partition function is holomorphically factorized as a function of the complex structure $\tau$ of the torus. The theories examined in \cite{threedqgadscftwitten} have an operator algebra that decomposes
completely into a tensor product of a right-moving and left-moving operator algebra, with no
further projections or additional sectors.
Any such theory must necessarily have a partition function that factorizes into a product of holomorphic and antiholomorphic factors, from which it follows directly that $\ccr,\ccl \in 24\IZ$.  In this class of CFT, it can be shown that the lowest primary operator is either purely left- or right-moving, and can have a weight no larger than $1 + {\rm min}({{\ccr}\over{24}},{{\ccl}\over{24}})$.  For all positive integer values of $({c\over{24}}, {{\tilde{c}}\over{24}})$, there exists a unique candidate partition function for which this bound is saturated, though it is not clear that this partition function necessarily corresponds to an actual conformal
field theory \cite{gab}.

%%%%%%%%%%%%%%%%%%%%%%%%%%%%%%%%%%%%%%%%%%%%%
%%%%%%%%%%%%%%%%%%%%%%%%%%%%%%%%%%%%%%%%%%%%%
\begin{figure}[htb]
\begin{center}
\includegraphics[width=5.0in,height=5.0in,angle=0]{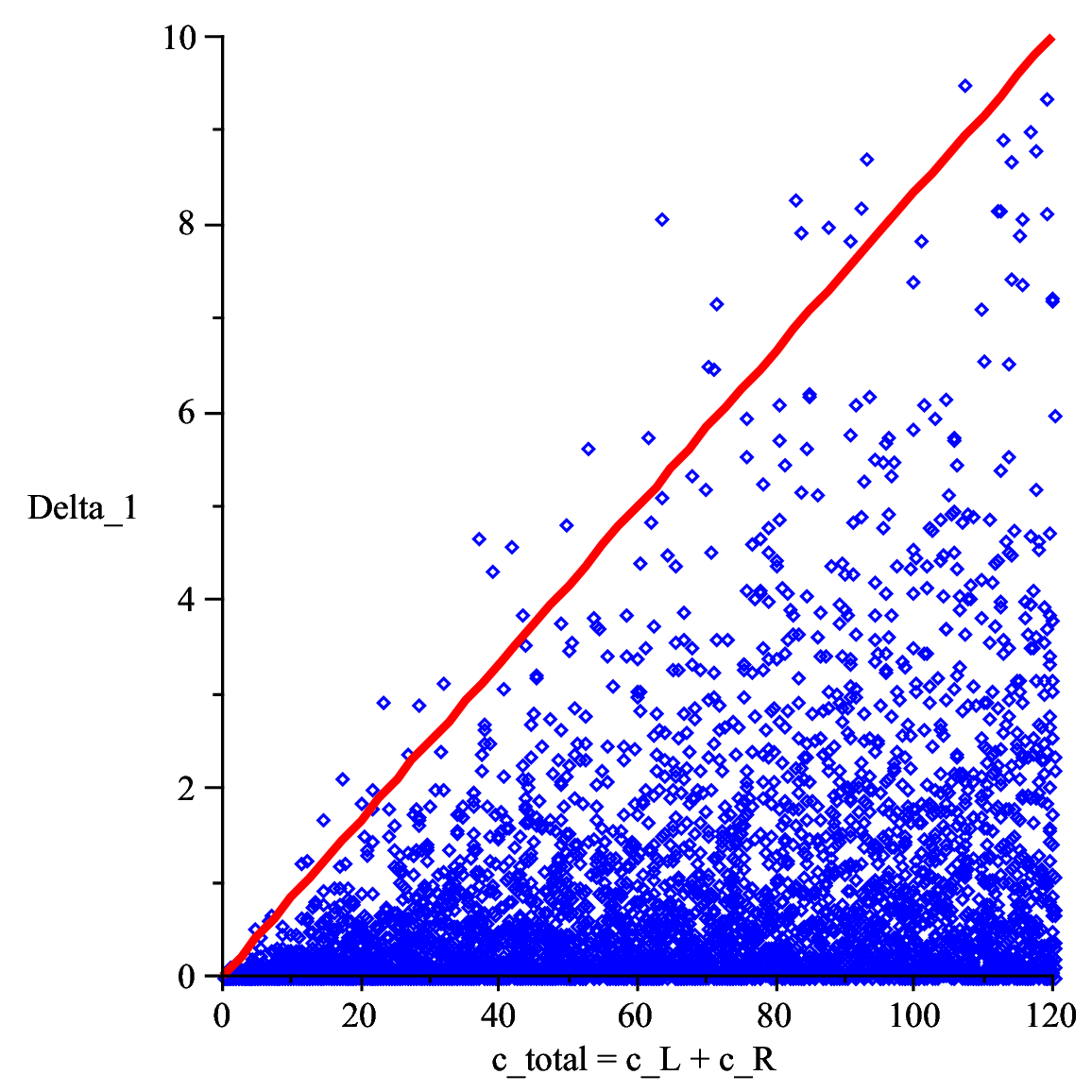}
\caption{One logical possibility is that there exists no "sharp" bound on
$\Delta_1$ / $\ctot$, but rather only a statistical falloff at large values of the ratio.}
\label{graph1}
\end{center}
\end{figure}
%%%%%%%%%%%%%%%%%%%%%%%%%%%%%%%%%%%%%%%%%%%%%
%%%%%%%%%%%%%%%%%%%%%%%%%%%%%%%%%%%%%%%%%%%%%

Other \subA{}{recent }work \cite{mooreetal} considers the case of theories with extended (2,2) supersymmetry, which allows the authors to exploit the power
of holomorphic dependence on the complex structure without assuming holomorphic factorization of the full partition function.  Study of a certain subclass of (2,2) SUSY CFT's suggests a bound that goes as $\D\ll 1 \leq {c\over{24}}$ for large central charge.  However it has not been possible so far to demonstrate this bound conclusively within this special class of SCFT nor to generalize the conjecture to all (2,2) SCFT, let alone to CFT with reduced supersymmetry or none at all.

In this paper we will derive a completely general lower bound on the weight
of the lowest primary operator in a completely general two dimensional conformal field theory, assuming only the basic properties of unitarity, modular invariance, and a discrete operator spectrum.\footnote{Even this last condition can be weakened substantially: Our conclusions will also apply to CFT with continuous spectrum that can be realized as limits of CFT with discrete spectrum.  The "singular points" in the moduli space of D1-D5 CFT
\cite{d1d51,d1d52}, for instance, are of this type \cite{seibwit, aspinwall, wittenk3}.}  Our bound will refer only to the energy of a \it single state, \rm namely the lowest excited primary state, rather than to the \edA{asymptotic} behavior of states at high \subA{energy}{temperature}.  Furthermore we will not use any sort of reasoning that refers to the bulk three-dimensional spacetime, nor make use of any asymptotic expansion at large central charge.  Our methods thus circumvent
the asymptoticity problem, and prove a general upper bound on the
lowest-weight primary in a general CFT.

%%%%%%%%%%%%%%%%%%%%%%%%%%%%%%%%%%%%%%%%%%%%%
%%%%%%%%%%%%%%%%%%%%%%%%%%%%%%%%%%%%%%%%%%%%%
\begin{figure}[htb]
\begin{center}
\includegraphics[width=5.0in,height=5.0in,angle=0]{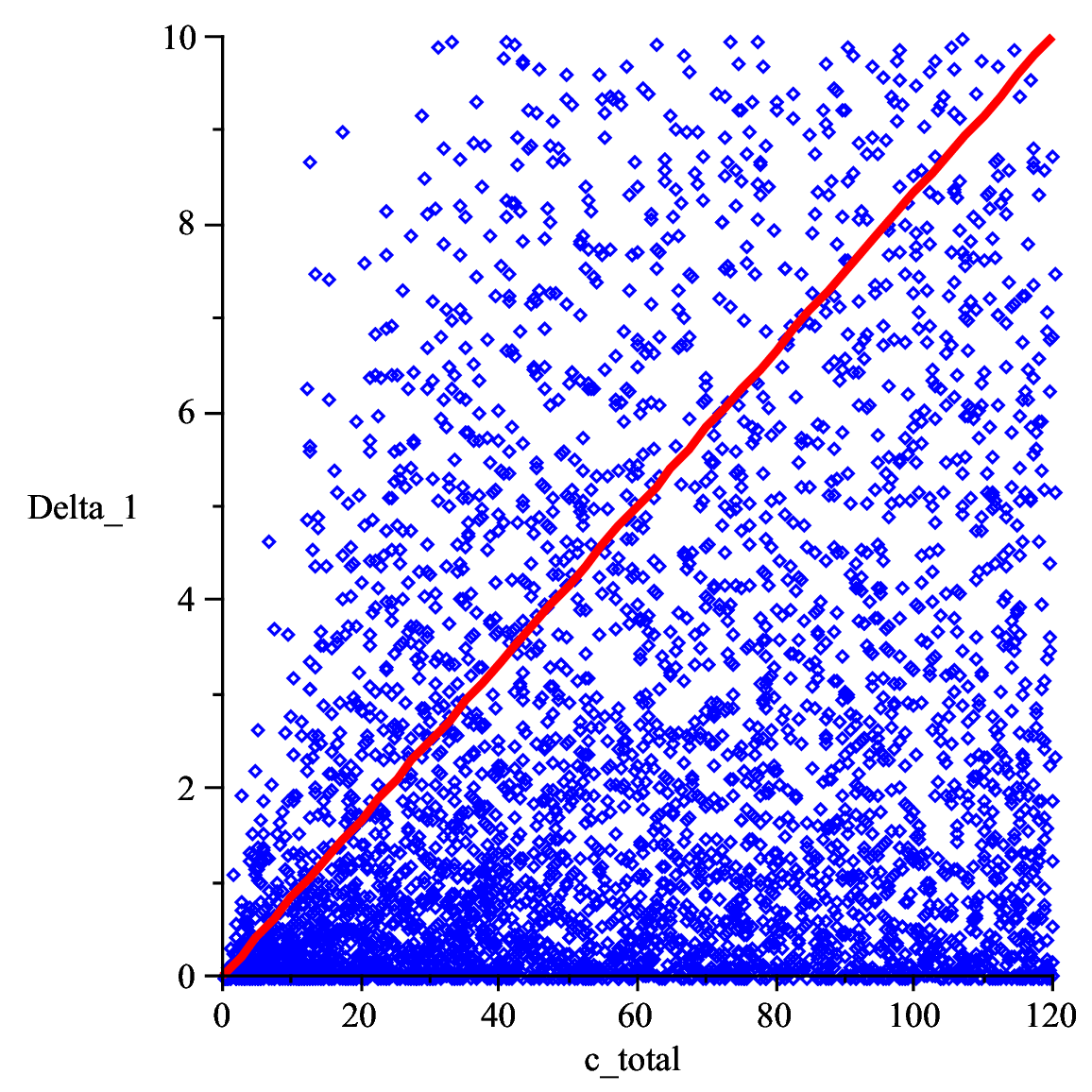}
\caption{One logical possibility is that there is no limit on $\Delta_1$ whatsoever for any given central charge, even in the "statistical" sense.}
\label{graph2}
\end{center}
\end{figure}
%%%%%%%%%%%%%%%%%%%%%%%%%%%%%%%%%%%%%%%%%%%%%
%%%%%%%%%%%%%%%%%%%%%%%%%%%%%%%%%%%%%%%%%%%%%

%%%%%%%%%%%%%%%%%%%%%%%%%%%%%%%%%%%%%%%%%%%%%
%%%%%%%%%%%%%%%%%%%%%%%%%%%%%%%%%%%%%%%%%%%%%
\begin{figure}[htb]
\begin{center}
\includegraphics[width=5.0in,height=5.0in,angle=0]{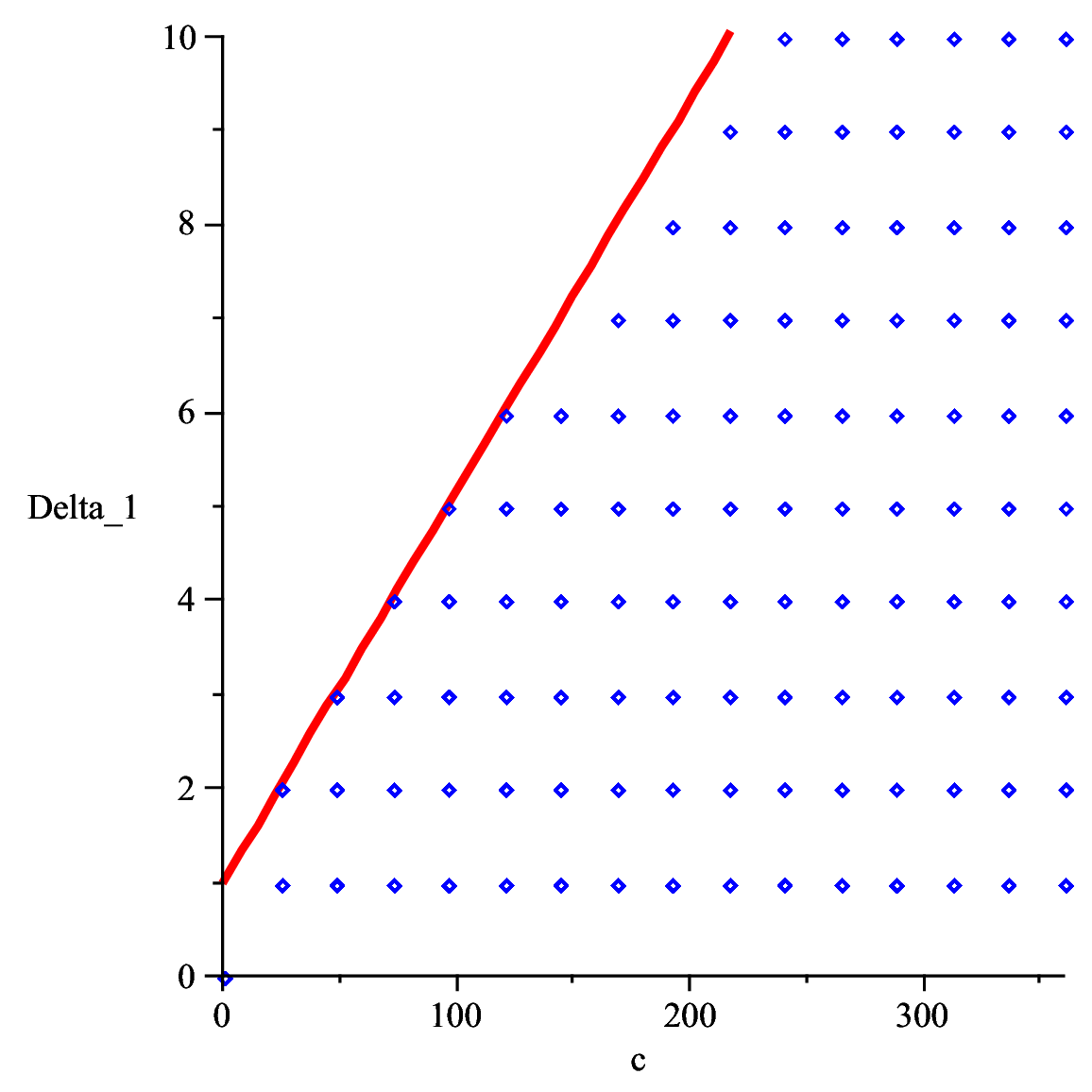}
\caption{In the case where the Hilbert space factorizes completely as a product of purely left- and right-moving CFTs, it is possible to show that $\Delta_1$ can never be greater than ${{\ctot}\over{24}} + 1$.  This is the case described by H\"ohn-Witten's conjectured "extremal" CFT.  It is unknown whether or not CFT exist that saturate this bound for $c$ equal to any positive integer multiple of 24.}
\label{graph3}
\end{center}
\end{figure}
%%%%%%%%%%%%%%%%%%%%%%%%%%%%%%%%%%%%%%%%%%%%%
%%%%%%%%%%%%%%%%%%%%%%%%%%%%%%%%%%%%%%%%%%%%%

 Our upper bound translates directly into an upper bound on the mass of the lightest massive state in a theory of gravity and matter in three dimensions.  The bound we derive applies to \it all \rm theories of gravity with an AdS$_3$ ground state.
   In particular, we do not assume
  holomorphic factorization, exact or approximate supersymmetry, or any other special property.\footnote{We will assume that the Hilbert space has a positive definite norm, and that the spectrum \edA{of} \subA{operator dimensions}{the Hamiltonian} is discrete.  Positivity is necessary for a consistent interpretation of quantum mechanics, and discreteness is necessary in order for the system to have well-behaved thermodynamic properties.  The assumption of discrete spectrum does not really count as a "special" property, as it holds on open sets of the moduli spaces of CFT that come in infinite families.}

\section{Inequalities from modular invariance}

In this section we will use unitarity and modular invariance to derive constraints on the energy levels of a general conformal field theory in two dimensions.  The techniques described in this section
were invented previously in \cite{cardy, cardy2}, where they were used to estimate dimensions of operators in special cases.
\footnote{In particular, see appendix A of \cite{cardy2}. We thank J. Cardy for correspondence relating to these and other results in the literature.}
More recently, related techniques have been used to bound certain operator
dimensions in conformal field theories in D=4\cite{rattazzi}.  (Also see \cite{seq1,seq2,seq3,seq4,seq5,seq6,seq7}.)  In this section,
we apply a similar method to bound the dimension
$\D\ll 1$ of the lowest primary operator in a general 2D CFT with $\ccl , \ccr < 9.135$.
This will serve as a warm-up to demonstrate the general ideas at work.  In the next
section we will derive an improved bound that applies to CFT with arbitrary central charges greater than 1.

\subsection{Conformal invariance and modular invariance}

Let us consider a general CFT in two dimensions with positive norm and discrete spectrum.  When the spatial direction $\s\uu 1$ of the theory is compactified on a circle of length $2\pi$, the partition function of the theory at temperature ${1\over \b}$ is given by
\bbb
Z(\b) \equiv {\rm Tr} \lrdd \exp{-\b H} \rrdd  = \sum\ll n a(n)\exp{ - \b E\ll n}\ ,
\eee
where $H$ is the Hamiltonian on a circle of length $2\pi$,
$E\ll n$ is the $n\uu{\underline{\rm th}}$ energy eigenvalue, and $a(n)$ is the degeneracy at
the energy $E\ll n$.
Unitarity and discreteness of the spectrum imply that the $a(n)$ are
positive integers.  (We will sometimes suppress the degeneracy $a(n)$ henceforth.)  The partition function can be refined by adding a thermodynamic
potential $K\uu 1$ for momentum $P\ll 1$ in the compact spatial direction
$\s\uu 1$:
\bbb
Z(\b, K\uu 1) \equiv {\rm Tr} \lrdd \exp{i K \uu 1 P\ll 1 - \b H}\rrdd
\eee
Defining $\t\equiv (K\uu 1 + i H) / 2\pi$, the partition function
can be written in the familiar form
\bbb
Z(\t,\tb) \equiv {\rm Tr} \lrdd q\uu{L\ll 0 - {{\ccr}\over{24}}} \bar{q}\uu
{\tilde{L}\ll 0 - {{\ccl}\over{24}}} \rrdd \ ,
\eee
where $q\equiv \exp{2\pi i \t}$,
$\ccr$ and $\ccl$ are the right- and left-moving central charges,
$L\ll 0 = \hh (H + P\ll 1) + {{\ccl}\over{24}}$ and $\tilde{L}\ll 0
= \hh (H - P\ll 1) + {{\ccl}\over{24}}$ are the right- and
left-moving conformal weight operators, and $q\equiv {2\pi i\t}$,
which fit into a Virasoro algebra
\bbb
\lsqq L\ll m, L\ll n \rsqq = (m - n) L\ll{m+n} + {{\ccr}\over{12}} (m\uu 3 - m) \d\ll{m,-n}
\xxx
 \lsqq \tilde{L}\ll m , \tilde{L}\ll n \rsqq = (m - n) \tilde{L}\ll{m+n} +
{{\ccl}\over{12}} (m\uu 3 - m) \d\ll{m,-n}
\eee
\bbb
[L\ll m, \tilde{L} \ll n] = 0
\eee
The Virasoro generators obey the Hermiticity condition $L\dag\ll m = L\ll{-m}$, and it follows
from unitarity that every primary operator has nonnegative weight, with weight zero if and
only if the operator is the identity.

The partition function can be realized as the path integral of
the conformal field theory on a torus of complex structure $\t$, with
no operator insertions.
Large coordinate transformations of the torus have the structure of the
modular group ${\bf PSL}(2,\IZ)$, with the generator $\lrdd \bm { a & b \cr c & d } \em \rrdd$
acting as $\t \to {{a \t + b}\over{c \t + d}}$.  The group is generated by the
transformations $T = \lrdd \bm { 1 & 1 \cr 0 & 1 } \em \rrdd$ and $S = \lrdd \bm {0 & -1\cr
1 & 0}\em\rrdd$, which act as $\t\to \t + 1$ and $\t \to - {1\over\t}$, respectively.

Invariance of the partition function under the $T$ transformation is completely equivalent to the condition that every state have $h - \tilde{h} \in \IZ$, where $h,\tilde{h}$ are the state's
eigenvalues under $L\ll 0, \tilde{L}\ll 0$.  By contrast, invariance of the partition function under the modular $S$ transformation gives a condition
that \edA{is} far less transparent as a set of constraints on the spectrum of the theory.
The rest of this section is
devoted to extracting useful information from the invariance of $Z(\t)$ under this transformation.

%\subsection{Analysis of the holomorphically factorized case}

\subsection{The medium-temperature expansion}

The most basic \subA{consequence of modular invariance}{constraint} is Cardy's formula \cite{cardy}, which has to
do with the asymptotic density of energy levels.  By relating the low-temperature limit $\t\ll 2 \to \infty$ directly to the high-temperature limit $\t\ll 2 \to 0$, the modular $S$-transformation implies that the asymptotic density of states goes as
\bbb
\rho(\D,J) \equiv {{d\sqd n}\over{d\D~dJ}} \simeq
\exp{\hilo 2\pi\lrdd \sqrt{{{\ccl (\D - J)}\over{12}}}
+ \sqrt{{{\ccr (\D + J)}\over{12}}} \rrdd }\ ,
\eee
where $\D = h + \tilde{h}$ is the scaling dimension of the operator, $J \equiv h - \tilde{h}$ is the spin of the operator, and we take $\D$ to be
large.  Setting to zero the chemical potential $K\uu 1$ for spin, we maximize the exponent with respect to $J$, which gives
\bbb
\rho(\D) \equiv
{{dn}\over{d \D}} \simeq \exp{ + 2\pi\sqrt{{\ctot\Delta}\over{6}}}
\eee
for large $\Delta$ at a given total central charge
\bbb
\ctot \equiv \ccr + \ccl\ .
\een{ctotdef}

Since the level density increases slower than any geometric progression, this
implies that the sum $\sum\ll n {\aaa n} (q\bar{q})\uu{E\ll n}$
converges for any $|q| < 1$, by
the ratio test.  This will be useful for us in what follows,
but by itself it does not make a
statement about the spectrum that can be tested at finite energies or temperatures -- Cardy's
formula is limited by its status as a formula that applies only asymptotically.

We would like to devise a test that allows us to look at a finite number of energy levels in a candidate spectrum for a CFT, and
to decide whether that set of energy levels can actually be the spectrum of a consistent theory.
And we'd like to do it by using as few inputs as possible -- in particular, we would like to see what can be accomplished just using unitarity and modular invariance, and not having to use other consistency constraints on CFT, such as the existence of an associative operator product expansion.

For instance, the considerations in \cite{ hohnone, threedqgadscftwitten}
yield a constraint on the spectrum of a CFT in the
case that the partition function of the CFT factorizes as a product of holomorphic and anti-holomorphic functions.  Using only unitarity, holomorphy and modular invariance, one can show that in a holomorphically factorized CFT, the weight of the lowest-lying primary
state (other than the identity) can be no higher than $\Delta\ll 1 \leq {\rm min}({{\ccl}\over{24}},  {{\ccr}\over{24}}) + 1$.

Similarly, in \cite{mooreetal}
the authors examined the case of CFT with $(2,2)$
superconformal invariance, and used the holomorphic properties of the
elliptic genus suggest a conjectural bound of $ \Delta\ll
 1 \leq  {1\over{24}} ~{\rm min}\lrdd \ccr, \ccl \rrdd  + o
\lrdd \ctot\uu 0 \rrdd  \rm$
at large central charge, under some special conditions.  (In
\cite{mooreetal}, the bound contains
more information than the
one in \cite{hohnone, threedqgadscftwitten}, in that the terms
subleading in $\ccl,\ccr$ have precise
coefficients depending on the $U(1)$ R-charge of the state of interest.)

To derive -- and fully prove -- a bound in the general case, let us conisder
the nature of the inputs we are using.  If we use only unitarity and modular invariance, what
we are really studying is just the set of modular invariant functions with a Fourier
expansion that is discrete, with positive integer Fourier coefficients.
What are the general properties of this set of functions?

Functions $f(\t,\tb)$ on the upper half plane $\IH$ that are modular invariant are in one to one correspondence with functions on the fundamental domain $\IH / {\bf PSL}(2,\IZ)$.
But if we further assume that $f(\t, \tb)$ is smooth on the covering space, this gives us
extra information, because the set of modular-invariant \it smooth \rm
functions $f(\t, \tb)$ is \it not \rm in correspondence with the set of smooth
functions on the fundamental region.
A smooth function on
the fundamental region lifts to a modular-invariant smooth function on the
covering space if and only if it satisfies certain conditions on its
derivatives at the special points on the fundamental region that correspond
to fixed points of elliptic elements of the modular group \cite{cardy,cardy2}.
Cardy's formula implies that $Z(\t, \bar{\t})$ and all its derivatives are continuous in
the entire upper half plane, so we can indeed apply this reasoning to the partition function.

We will focus
here on the point $\t = +i$, which is fixed under the modular
transformation $S = \lrdd \bm { 0 & -1 \cr 1 & 0 } \em \rrdd $.  This
point in complex structure moduli space corresponds to a torus
that is square -- that is, the metric on
the torus has no shear (off-diagonal components), and both radii are equal.
The path integral on a square torus corresponds to the thermal partition
function of a CFT compactified on a circle, at a temperature equal to the
inverse circumference of the circle.  The partition function at
higher temperatures can be expressed, using the modular $S$-transformation,
in terms of the partition function at lower temperatures.  So the complex
structure $\t = +i$, corresponding to $\b = {1\over {k\ll {\rm B}{\rm T}}}
= 2\pi$, can be thought of as lying exactly between the high-temperature
and low-temperature r\'egimes, or equivalently between
the large- and small-complex structure limits of the moduli space of
the torus.

It is in the neighborhood of $\t = i$ that modular invariance of the
parition function, which is obscure in the large- and small-complex structure
limits, becomes manifest.  To parametrize this neighborhood conveniently,
write
\bbb
\t \equiv + i \cdot \exp{s}\ .
\eee
In terms of the variable $s$, the modular $S$-transformation $\t\to -
{1\over \t}$ acts as
\bbb
s\to - s\ .
\eee

In terms of the variable $s$, then, the condition of invariance under
the modular $S$-transformation can be written as
\bbb
Z(+i ~ \exp{s}, - i ~ \exp{\bar{s}}) = Z(+i ~ \exp{-s}, - i ~
\exp{-\bar{s}})\ .
\eee
Scaling $s \to 0$ and examining the behavior of the partition function
in that r\'egime is what we shall refer to as the \it medium complex
structure expansion, \rm or \it medium temperature expansion \rm when we restrict
ourselves to real values of $s$.

Taking derivatives at $s = 0$, this gives
\bbb
\lno\hilo \lrdd {{\pp}\over{\pp s}} \rrdd \uu{N\ll {\rm R}}
\lrdd {{\pp}\over{\pp \bar{s}}} \rrdd \uu{N\ll {\rm L}} Z(+i ~ \exp{s},
- i ~ \exp{\bar{s}})
\rba\ll{s = 0} = 0 ~{\rm for}~N\ll{\rm R} + N\ll {\rm L}~{\rm odd}\ .
\eee
In terms of the usual variable $\t$, this means
\bbb
\lno\hilo \lrdd \t {{\pp}\over{\pp \t}} \rrdd \uu{N\ll {\rm R}}
\lrdd \tb { {\pp}\over{\pp \tb}} \rrdd \uu{N\ll {\rm L}} Z(\t,
\tb)
\rba\ll{\t = +i} = 0 ~{\rm for}~N\ll{\rm R} + N\ll {\rm L}~{\rm odd}\ .
\eee

For purely imaginary complex structures $\t = i \b / 2\pi$, this
condition implies
\bbb
\lno \lrdd \b {{\pp}\over{\pp\b}} \rrdd\uu N Z(\b)
\rba\ll{\b = 2\pi} = 0~{\rm for}~N~{\rm odd}\ .
\eee

Before proceeding to derive our inequalities, let us make a few remarks
on the medium temperature expansion.  First of all, we want to emphasize
that this expansion really is quite powerful -- the $N\uu{\underline{\rm
th}}$ order of the medium temperature expansion generates a new
constraint on the partition function, indepdendent from all the
previous ones.

Secondly, the medium temperature expansion really contains
complementary information to that of the low- and high-temperature limits.
In the low-temperature r\'egime, unitarity is
manifest because the function decomposes into a sum of exponentials with
positive integer coefficients, but modular invariance is completely
invisible.  The high-temperature r\'egime is redundant with the
low-temperature r\'egime, and Cardy's
formula exploits this relation in order to derive asymptotic formul\ae
for level densities.  But still modular invariance constrains neither
of these two limits separately, it only relates them to one another.  In
the medium temperature r\'egime, unitarity is obscure, but
modular invariance is manifest and imposes
an infinite number of separate constraints on the derivatives
of the partition function.

Thirdly, we will see in the next section
that the medium temperature expansion is \it useful, \rm when combined
with the unitarity constraints that are visible in the low-temperature
expansion and the convergence properties inferred \it via \rm Cardy's formula
for the high-temperature expansion.
We will see that one need not use the full hierarchy of differential
constraints on $Z(\b)$ for all odd $N$: there is a
useful inequality implied just by the
constraints at $N=1,3$ alone.  We now turn to the derivation of this
inequality.

\subsection{Warm-up : the case of $\ccr, \ccl < 9.135$}

In this subsection we will perform a ``warm-up'' derivation, where we
show that every compact, unitary CFT (with \subA{$\ccr,\ccl > 1$}{$c>1$}) has a local
operator of some kind -- not necessarily a primary operator --
whose scaling dimension $\D\equiv h + \tilde{h}$ satisfies
\bbb
\D \leq \D\ll +\uwu \equiv {{\ctot}\over{12}} + {3\over{2\pi}}\ .
\een{firstbound}
Since we include descendants in our partition function along with
primary operators, the bound \rr{firstbound} gives us interesting information only for low values of the total central charge:
for $\ctot  > 24 - {{36}\over{\pi}} \simeq 18.270$
the right hand side of \rr{firstbound} lies above $\D = 2$, where there must always be a stress tensor anyway.  When $\ctot < 18.270$, then $\D\ll +
\uwu < 2$, which means that the operator with dimension $\D\ll 1$ cannot
be a descendent of the identity.  It must either be primary, or else it
must be the $L\ll {-1}$ or $\tilde{L}\ll{-1}$ descendent of a primary
with dimension $\D\ll 1 - 1$.  In either case, this means there is a
nontrivial primary with dimension less than or equal to $\D\ll 1 < \D\ll +$.

Our method for deriving the inequality can be
refined to deal with primary operators \subA{specifically}{only}, and we will do this later
in the following sections.  For now, we will derive our weaker bound in order to illustrate the basic method involved.

First, we take the expression for the full partition function
\bbb
Z(\b) = \sum\ll n ~ {\aaa n} ~ \exp{- \b E\ll n}
\eee
and decompose it as
\bbb
Z(\b) = Z\upp{\rm vac} (\b) +  Z\upp{\rm excited} (\b)
\een{decomposition}
where
\bbb
Z\upp{\rm vac} (\b) \equiv \exp{- \b E\ll 0}
\xxn
Z\upp{\rm excited} (\b) \equiv \sum\ll {n \geq 1} ~ {\aaa n} ~  \exp{- \b E\ll n}\ .
\een{highlowdefs}

The basic idea behind the inequality we want to derive is to show
that with every derivative $\b{{\pp}\over{\pp \b}}$ that we take
at $\b = 2\pi$, the higher enegy levels tend to gain in
importance and also to contribute negatively to derivatives of odd order.  If we assume the degeneracies at high
levels are large enough that the first derivative with respect
to $\b$ will vanish at $\b = 2\pi$, then those high levels will tend
to contribute \it even more negatively \rm to the third derivative at $\b = 2\pi$, and the only way they can be balanced out
is by the lower positive energy levels.

So the conclusion will be that the lowest of the
excited energy level must be
low enough that it does not make an overwhelmingly large negative contribution
to the third derivative, given that it makes a \it sufficiently large \rm
contribution to the first derivative.  Let us now make this intuition
precise.  We define a \it relative importance \rm $I(E)$ associated with each energy level
$\kket{E}$.  This function is designed to
measure the importance of
$\kket{E}$'s contribution to the third derivative of the partition function at
medium temperature
($\b = 2\pi$), compared to $\kket{E}$'s
contribution to the first derivative at medium temperature.  The energy level $\kket{E}$
contributes to $Z(\b)$ as $\exp{- \b E}$ at temperature $1 / \b$, so we define
\bbb
I(E) ~~  \equiv  ~~
\lno
{
{
\bm{
\lrdd \b {{\pp}\over{\pp\b}} \rrdd \uu 3 \exp{- \b E}
 \cr {}
}\em
\over{
      \bm{
           {}
              \cr
                 \lrdd \b {{\pp}\over{\pp\b}} \rrdd \uu 1    \exp{- \b E}
          }\em }
}
}
\rba\ll{\b = 2\pi}
\xxn
~~  =  ~~ 4\pi\sqd E\sqd - 6 \pi E + 1
\een{ifuncdef}

At $\b = 2\pi$, the
derivatives of $Z\upp{\rm vac}$ and $Z\upp{\rm excited}$
will be equal and opposite, by virtue of
the medium-temperature expansion at $N=1$:
\bbb
\lno \lrdd \b {{\pp}\over{\pp\b}}\rrdd\uu 1
 Z\upp{\rm vac} (\b) \rba \ll{\b = 2\pi}
= \lno - \lrdd \b {{\pp}\over{\pp\b}} \rrdd\uu 1
Z\upp{\rm excited} (\b)\rba\ll{\b = 2\pi} \ .
\een{firstorder}
Likewise by virtue of the medium-temperature expansion at order $N=3$, we
have:
\bbb
\lno \lrdd \b {{\pp}\over{\pp\b}}\rrdd\uu 3
 Z\upp{\rm vac} (\b) \rba \ll{\b = 2\pi}
= \lno - \lrdd \b {{\pp}\over{\pp\b}} \rrdd\uu 3
Z\upp{\rm excited} (\b)\rba\ll{\b = 2\pi} \ .
\een{thirdorder}
(The zero energy states make no contribution to any of the derivatives.)

Now define the ratios
\bbb
{\cal R}\ll{31}\upp{\rm vac}
~~  \equiv  ~~
\lno
{
{
\bm{
\lrdd \b {{\pp}\over{\pp\b}} \rrdd \uu 3 Z\upp{\rm vac} (\b)
 \cr {}
}\em
\over{
      \bm{
           {}
              \cr
                 \lrdd \b {{\pp}\over{\pp\b}} \rrdd \uu 1   Z\upp{\rm vac}(\b)
          }\em }
}
}
\rba\ll{\b = 2\pi}
= I(E\ll 0)
\eee
and
\bbb
{\cal R}\ll{31}\upp{\rm excited}
~~  \equiv  ~~
\lno
{
{
\bm{
\lrdd \b {{\pp}\over{\pp\b}} \rrdd \uu 3 Z\upp{\rm high} (\b)
 \cr {}
}\em
\over{
      \bm{
           {}
              \cr
                 \lrdd \b {{\pp}\over{\pp\b}} \rrdd \uu 1   Z\upp{\rm high}(\b)
          }\em }
}
}
\rba\ll{\b = 2\pi}
\een{calrdefs}
The ratio ${\cal R}\ll{31}\upp{\rm vac}$ is equal to $I(E\ll 0)$
because the partition function $Z\upp{\rm vac}(\b)$ contains only a single exponential $\exp{- \b E\ll 0}$.

The ratios \rr{calrdefs} are necessarily equal in any modular invariant theory, by virtue of
relations \rr{firstorder} and \rr{thirdorder}:
\bbb
{\cal R}\ll{31}\upp{\rm excited} = {\cal R}\ll{31}\upp{\rm vac} = I(E\ll 0)\ .
\een{necessary}

Now we will use the relative importance factor $I(E)$ to show that \rr{necessary} can never
be satisfied if the bound \rr{firstbound} is violated.

To do this, write the ratio ${\cal R}\ll{31}\upp{\rm excited}$ as follows:
\bbb
{\cal R}\ll{31}\upp{\rm excited} = {{\sum\ll{m = 1}\uu\infty  I(E\ll m) {\aaa m} E\ll m \exp{- 2\pi E\ll m}}\over
{\sum\ll{n = 1}\uu\infty {\aaa n}  E\ll n \exp{- 2\pi E\ll n}}}\ .
\een{calrexpr}

Now let us compare the individual factors $I(E\ll n)$ to the ratio $I(E\ll 0) = {\cal R}\ll{31}\upp{\rm vac}$.  The equation $I(E) = I(E\ll 0)$ has two roots, namely $E\ll 0$ itself, and
\bbb
E\ll + \uwu \equiv {3\over{2\pi}} - E\ll 0 = {3\over{2\pi}} + {{\ccr + \ccl}\over{24}}\ .
\een{eplusdef}
The larger root $E\ll +\uwu$ is positive, since $E\ll 0$ is negative in any unitary theory.
So there are two possible ranges for the relative importance function $I(E)$, namely
\bbb
I(E) \leq (E\ll 0) ~~~~~~~~~~~~~~ {\rm for}~~~ E\ll 0 \leq E \leq E\ll +\uwu
\een{saferange}
and
\bbb
I(E) > (E\ll 0) ~~~~~~~~~~~~~~ {\rm for}~~~ E > E\ll +\uwu \ .
\een{badrange}

Now we will use proof by contradiction to show that the lowest excited energy level $E\ll 1$ can be no greater than $E\ll +$.  Suppose $E\ll 1$ lies in the second range, \rr{badrange}.  Then
so does every excited level $E\ll n, n\geq 1$.  This would give us the inequalities
\bbb
E\ll n \geq E\ll 1 > E\ll +\uwu > 0\ ,
\xxn
I(E\ll n) > I(E\ll 0) > 0
\een{inequalitiesone}
for all $n \geq 1$.   These inequalities lead to a contradiction: subtracting the two sides of \rr{necessary} and
using the identity \rr{calrexpr}, we obtain
\bbb
0 = {\cal R}\ll{31}\upp{\rm excited} - I(E\ll 0)
=  {{\lno \sum\ll{m = 1}\uu\infty
\hilo \lrdd I(E\ll m) - I(E\ll 0) \rrdd \cdot E\ll m \cdot {\aaa m} ~ \exp{- 2\pi E\ll m} \rno}
\over
{\lno \sum\ll{n = 1}\uu\infty
\hilo E\ll n \cdot {\aaa n} ~ \exp{- 2\pi E\ll n} \rno}}\ .
\een{setupone}
If $E\ll 1 > E\ll +\uwu $, then
every term in the numerator and denominator of \rr{setupone} is positive, by virtue of the inequalities \rr{inequalitiesone}, and equation \rr{setupone} cannot be satisfied.
We conclude that
\bbb
E\ll 1 \leq {{\ctot }\over{24}} + {3\over{2\pi}} \ , ~~~~~~~~~~~~~~~~~~~~~~~~{\it Q.E.D.}
\een{qedone}

The proof above establishes a general upper bound for the lowest excited energy level
in any unitary CFT with discrete spectrum.  Written in terms of operator dimensions $\Delta
\equiv E - E\ll 0$, we have a lower bound on the scaling dimension of the lowest-dimension operator other than the identity:
\bbb
\Delta\ll 1 \leq \Delta\ll +\uwu \ ,
\een{warmupresult}
with $\Delta\ll + \uwu \equiv {{\ctot}\over{12}} + {3\over{2\pi}}$.  For $\ccr + \ccl \leq 24 - {{18}\over{\pi}} \simeq 18.270$,
we have $\D\ll +\uwu < 2$, so the lowest operator satisfying the bound, other than the identity, must be a primary operator.  For $\ctot > 18.270$, the value of $\D\ll +\uwu $ is greater than 2, so the bound yields no information, since any CFT always contains a stress tensor,
which has dimension two.

In the range $\ctot \leq 18.270$, the individual central
charges $\ccr$ and $\ccl$ must be equal, since both are positive (by unitarity) and their
difference is an integer multiple of 24 (by modular invariance).  So the
theorem states that there exists a primary operator (other than the identity) of scaling dimension less than ${{c}\over 6} + {3\over{2\pi}}$ for any unitary, modular invariant CFT with $ \ccr,\ccl < 9.135$.
This is a somewhat limited range of central charge, but it contains many interesting theories,
including any supersymmetric sigma model on a Calabi-Yau threefold (with diagonal GSO projection).

\section{A general inequality for primary operators}

In order to derive a useful inequality for theories with $\ctot > 18.270$, we would like to separate primaries from descendants, so that we can find an upper bound on the weight of the lowest \it primary \rm state.  In this section, we will adapt the methods of the previous section
to focus on primary operators alone.

\subsection{Strategy}

Our strategy to
derive such an upper bound, in parallel with the derivation of the previous section, is to write the partition function explicitly in terms of weights
of primaries, rather than in terms of weights of general operators.
To do this, we proceed as follows:

\begin{enumerate}

\item{Restrict our attention to the case where both left- and right-moving central charges are greater than 1.  In this case, the Virasoro representations of primary states are particularly simple.  We will simplify the analysis further by assuming that the CFT has no chiral algebra other than the Virasoro algebra.}

        \item{Write down \edA{the partition function of the full theory in terms of }a partition function for primary operators alone.  For the cases we consider,\subA{ these two functions are related by a simple linear algebraic equation.}{ this partition function is related by a simple equation to the full partition function.}}

            \item{Use the relation between the full partition function and the partition function for primaries to derive \subA{an identity }{a transformation law }for the latter under the modular $S$-operation $\t\to - {1\over \t}$, given the modular invariance of the former.}

                \item{Express modular invariance as an infinite sequence of \subA{linear identities among derivatives of}{differential identities on} the partition function for primaries at medium temperature $\b\uu{-1} = {1\over{2\pi}}$.}

                    \item{Show that the first two of these identities are not compatible with one another if the energy of the lowest primary state is too large.}

                        \end{enumerate}

\subsection{Review of Virasoro representations}

For the rest of this section we will assume $\ccr$ and $\ccl$ are both greater
than $1$.  This has well-known and useful implications for the structure of representations
of the Virasoro algebra \cite{kacdet, feiginfuks, details}.  In particular, for $c > 1$
the unitary highest-weight representations of the Virasoro algebra are of two types,
characterized by the weight $h$ of the primary state $\kket{h}$ on which the representation
is built.  For the first type of representation, the primary state $\kket{h}$ has weight $h>0$, and all each ordered monomial of Virasoro raising operators creates an independent
state.  That is, for $h > 0$ there are no linear relations among the states
\bbb
L\ll{-n\ll 1} \cdot L\ll{- n\ll 2} \cdots L\ll{- n\ll k} \cdot \kket{h}
\eee
for any collection (possibly empty) of $n\ll i$ with $n\ll 1 \geq n\ll 2 \geq \cdots \geq n\ll k \geq 1$.  The second type of Virasoro representation is the one in which the primary state
has $h = 0$.  In this case the linearly independent states of the representation are given by
\bbb
L\ll{-n\ll 1} \cdot L\ll{- n\ll 2} \cdots L\ll{- n\ll k} \cdot \kket{0}
\eee
for any collection (possibly empty) of $n\ll i$ with $n\ll 1 \geq n\ll 2 \geq \cdots \geq n\ll k \geq 2$.

\subsection{Decomposition of the partition function}

By assuming the theory has no chiral algebra beyond the Virasoro algebra, we eliminate from
consideration primaries with $h = 0, \tilde{h} \neq 0$ or vice versa.  So the only primaries
in our theory have $h = \tilde{h} = 0$ or $h,\tilde{h} > 0$.  By cluster decomposition, the CFT can contain only one operator with $h = \tilde{h} = 0$, namely the identity operator.  So we can decompose our partition function $Z(\t)$ into a sum over conformal families, including the
identity family:
\bbb
Z(\t) = Z\ll {\rm id }(\t) + \sum\ll A Z\ll A(\t) \ ,
\eee
where $Z\ll {\rm id}(\t)$ is the sum over states in the conformal family of the identity, and
$Z\ll A(\t)$ is the sum over all states in the conformal family of the $A\uu{\rm {\underline{th}}}$ primary, which has conformal weights $h\ll A, \tilde{h}\ll A$.  By the
structure theorem for Virasoro representations with $c > 1$ referred to above \cite{kacdet, feiginfuks, details}, the partition function $Z\ll A(\t)$ is
\bbb
Z\ll A (\t)
= q\uu{h\ll A -{\ccr\over{24}}} \bar{q}\uu{\tilde{h}\ll A-{{\ccl}\over{24}}}
\prod\ll{m = 1}\uu\infty
\lrdd 1 - q\uu m \rrdd\uu{-1} \prod\ll{n = 1} \uu\infty
\lrdd 1 - \bar{q}\uu n \rrdd\uu{-1}
\eee

Likewise, the partition function $Z\ll {\rm id} (\t)$ is
\bbb
Z\ll{\rm id}(\t) = q\uu{ -{\ccr\over{24}}} \bar{q}\uu{-{{\ccl}\over{24}}}
\prod\ll{m = 2}\uu\infty \lrdd 1 - q\uu m \rrdd\uu{-1} \prod\ll{n = 2}\uu\infty \lrdd 1 - \bar{q}\uu n \rrdd\uu{-1}
\eee

So we can write the full partition function as
\bbb
Z(\t) = q\uu{- {{\ccr}\over{24}}} \bar{q}\uu{- {{\ccl}\over{24}} }
\lsqq \prod\ll{m = 1}\uu\infty \lrdd 1 - q\uu m \rrdd\uu{-1} \rsqq\cdot
\lsqq \prod\ll{n = 1}\uu\infty \lrdd 1 - \bar{q}\uu n \rrdd\uu{-1} \rsqq\cdot
\lsqq (1 - q)(1 - \bar{q}) + Y(\t) \rsqq
\eee
where
\bbb
Y(\t) \equiv \sum\ll {A=1}\uu\infty q\uu{-h\ll A} \bar{q}\uu{- \tilde{h}\ll A}
\eee
is a sum over primary states only, with the vacuum omitted.

We can simplify this expression by using the definition of the Dedekind eta function
\bbb
\eta(\t) \equiv q\uu{+{1\over{24}}} \cdot \prod\ll{n = 1}\uu\infty \lrdd 1 - q\uu n \rrdd\uu{+1}
\eee
so
\bbb
\prod\ll{n = 1}\uu\infty \lrdd 1 - q\uu n \rrdd\uu{-1} = {{q\uu{+{1\over{24}}}}\over{\eta(\t)}}\ .
\eee
So we write the full partition function as
\bbb
Z(\t) = q\uu{- {{\ccr - 1}\over{24}}} \bar{q}\uu{- {{\ccl - 1}\over{24}}}\cdot
 |\eta(\t)|\uu{-2} \cdot
 \lsqq (1 - q)(1 - \bar{q}) + Y(\t) \rsqq
\eee
Restricting to the imaginary axis $\t \equiv i \b / (2\pi)$, with $\b$ real, we have $q = \bar{q} =
\exp{-\b}$, and
\bbb
Z(\b) = M(\b) Y(\b) + B(\b)
\een{zfac}
with
\bbb
M(\b) \equiv {{\exp{-  (E\ll 0 + {1\over{12}})\b}}\over{\eta(i \b / 2\pi )\sqd}}
\een{mdef}
and
\bbb
B(\b) \equiv M(\b) \cdot \lrdd 1 - \exp{-  \b} \rrdd \uu{+2}\ ,
\een{bdef}
where $E\ll 0 \equiv -{{\ctot }\over{24}}$.  For real $\b$, the partition function over
primaries $Y(\b)$ becomes
\bbb
Y(\b) \equiv \sum\ll A \exp{- \b \Delta \ll A}\ ,
\een{ydef}
where $\Delta\ll A \equiv h_A + \tilde{h}_A$
is the weight of the primary operator $\co\ll A$.

In what follows, it will be convenient to define \subA{a new set of}{the} polynomials $f\ll p(x)$ by the equation
\bbb
\lno\hilo \lrdd \b\pp\ll\b \rrdd\uu p \lsqq {{\exp{-z\b}}\over{\eta(i\b / 2\pi)\sqd}} \rsqq
\rba\ll{\b = 2\pi}
\equiv  (-1)\uu p \cdot \eta(i)\uu{-2} \cdot \exp{-2\pi z}  \cdot f\ll p (z)\ .
\een{formula}
The first few polynomials are
\bbb
f\ll 0 (z) = 1
\xxn
f\ll 1(z)  = (2\pi z) - \hh
\xxn
f\ll 2 (z) = (2\pi z)\sqd -2\, (2\pi z) + \lrdd  {7\over 8} + 2 r\ll{20} \rrdd
\xxn
f\ll 3 (z) = (2\pi z)^3 - {9\over 2} (2\pi z)^2 + \lrdd {{41}\over 8} + 6 r\ll{20} \rrdd (2\pi z) - \lrdd {{17}\over{16}}
+ 3 r\ll {20} \rrdd
\een{defpolys}
where $r\ll{20}$ is a numerical constant we have defined as
\bbb
r\ll {20} \equiv {{\eta\uu{\prime\prime}(i)}\over{\eta(i)}} = \lrdd {{\eta\pr(i)}\over{\eta(i)}}
\rrdd\sqd + \pp\ll\t\sqd \lno \lsqq {\rm ln}\lrdd \eta(\t) \rrdd\rsqq \rba\ll{\t = i}
= - {1\over{16}} +  \sum\ll{n = 1}\uu\infty {{\pi\sqd n\sqd}\over{ {\rm sinh}\sqd (\pi n)}}
\xxn
= 0.0120528 + o\lrdd 10\uu{-8} \rrdd\ .
\een{rtwozerodef}
In deriving these polynomials we have used the identities
\bbb
\eta\pr(i) = {i\over 4} \eta(i)
\xxn
\eta\uu{\prime\prime\prime}(i) = {{15 i }\over {32}} \lrdd \eta(i) + 8 \eta\uu{\prime\prime}(i) \rrdd \ .
\een{etaidents}
The identities \rr{etaidents} follow from the modular transformation law of the eta function,
\bbb
\eta(- {1\over \t}) = (-i \t)\uu\hh \cdot \eta(\t)\ ,
\eee
whose medium-complex-structure expansion yields
\bbb
\lno\hilo  \lrdd \t\pp\ll\t + {1\over{4}} \rrdd\uu p \eta(i) \rba\ll{\t = i} = 0,~~~
p{~\rm odd}\ .
\eee
The $p=1,3$ identities yield eqns. \rr{etaidents} directly.

Now we would like to take derivatives of the two terms $B(\b)$ and $M(\b) Y(\b)$ at
medium temperature.  For $M(\b) Y(\b)$, equations \rr{mdef},\rr{ydef} together with
the formula \rr{formula} give
\bbb
\lno\hilo \lrdd \b\pp\ll\b \rrdd\uu p M(\b) Y(\b) \rba\ll{\b = 2\pi} =
\xxn
 \llsk\llsk (-1)\uu p
\eta(i)\uu{-2} \exp{- 2\pi(E\ll 0 + {1\over{12}})} \sum\ll{A = 1}\uu\infty f\ll p
\lrdd \Delta\ll A
+ E\ll 0 + {1\over{12}}\rrdd \exp{- 2\pi\Delta\ll A}
\een{myderivs}
As for $B(\b)$, applying the formula \rr{formula} to eq. \rr{bdef} gives
\bbb
\lno\hilo \lrdd \b\pp\ll\b \rrdd\uu p B(\b) \rba\ll{\b = 2\pi} =
(-1)\uu p
\eta(i)\uu{-2} \exp{- 2\pi(E\ll 0 + {1\over{12}})} \cdot
\xxn
\lsqq
 f\ll p\lrdd E\ll 0 + {1\over{12}}\rrdd - 2 \, \exp{- 2\pi}\,
 f\ll p\lrdd E\ll 0 + {{13}\over{12}} \rrdd
 + \exp{- 4\pi}\, f\ll p \lrdd E\ll 0 + {{25}\over{12}} \rrdd \rsqq
\een{bigBderivs}
\def\ehh{\hat{E}_0}
For simplicity, we now define $\ehh \equiv E\ll 0 + {1\over{12}} = {{2 - \ctot}\over{24}}$.
Then we have
\bbb
\lno\hilo \lrdd \b\pp\ll\b \rrdd\uu p M(\b) Y(\b) \rba\ll{\b = 2\pi} =
(-1)\uu p
\eta(i)\uu{-2} \exp{- 2\pi \ehh  } \sum\ll{A = 1}\uu\infty f\ll p(\D\ll A + \ehh) \exp{- 2\pi\Delta\ll A}
\xxn
\lno \hilo \lrdd \b\pp\ll\b \rrdd\uu p B(\b) \rba\ll{\b = 2\pi} =
(-1)\uu p
\eta(i)\uu{-2} \exp{- 2\pi\ehh } \cdot b\ll p (\ehh)
\xxn
b\ll p (x) \equiv
f\ll p (x ) - 2\, \exp{- 2\pi} f\ll p (x +  1) + \exp{- 4\pi} f\ll p ( x + 2)
\een{simpderivs}
Thus the medium-temperature equations for modular invariance of $Z(\b)$ at $p = 1,3$ reduce to
\bbb
\sum\ll {A = 1}\uu\infty f\ll 1 (\D\ll A + \ehh) \, \exp{- 2\pi \D\ll A} = - b\ll 1 (\ehh)
\xxn
{\rm and}
\xxn
\sum\ll {A = 1}\uu\infty f\ll 3 (\D\ll A + \ehh) \, \exp{- 2\pi \D\ll A} = - b\ll 3 (\ehh)
\een{modularprim}
Now we will proceed in parallel with our "warm-up" proof.  Define the relative importance
factor
\bbb
I\ll{31}(x) \equiv {{f\ll 3(x)}\over{f\ll 1(x)}}
\een{iprimdef}
and the coefficient
\bbb
K\ll{31}(\ehh) \equiv {{b\ll 3 (\ehh)}\over{b\ll 1 (\ehh) }}\ .
\een{kdef}
As before, we will take the scaling dimensions $\D\ll n$ of our primary operators to be indexed
in order so that $\D\ll n$ is increasing:
\bbb
0  = \D\ll 0  < \D\ll 1 \leq \D\ll 2 \leq \cdots \ .
\eee
 To derive an upper bound on the dimension $\D\ll 1$ of the lowest primary operator other than the identity, divide the two equations \rr{modularprim}.  We then obtain
\bbb
 {{ \sum\ll {A= 1}\uu\infty I\ll {31}(\D\ll A + \ehh) f\ll 1(\D\ll A + \ehh) \exp{- 2\pi \Delta\ll A}}\over
 { \sum\ll {B= 1}\uu\infty ~~~~~~~~~~~~ f\ll 1(\D\ll B + \ehh ) \exp{- 2\pi \Delta\ll B} } } = K\ll{31} (\ehh)
\een{necessaryprim}
Now we will prove that (for $\ccr,\ccl > 1$) the value of $\D\ll 1$ must always be less than
or equal to $\D\ll +$, which we define as the largest \edA{real }solution $\D$ to the equation
\bbb
I\ll{31}(\D + \ehh ) = K\ll{31} (\ehh)\ ,
\een{firstversionxplus}
which is equivalent to the cubic equation
\bbb
f\ll 3(\D + \ehh ) - K\ll{31}(\ehh)  f\ll 1 (\D + \ehh) = 0\ .
\een{secondversionxplus}
% EDITING DONE UP TO HEER ON MARCH 22
Note that $\D\ll + $ is implicitly a function of $\ehh$,
though we will not always indicate the $\ehh$ dependence in our notation.

We will prove our desired result by contradiction.  Suppose $\D\ll 1 > \D\ll +$.  Now subtract the two sides of
\rr{necessary}, to give
\bbb
0 = {{\sum\ll{A = 1}\uu\infty
 \lrdd I\ll{31}(\D\ll A + \ehh) - K\ll{31}(\ehh) \rrdd \cdot f\ll 1 (\D\ll A + \ehh)\cdot
\exp{- 2\pi \D\ll A}}\over{\sum\ll{B = 1}\uu\infty ~~~~~~~~~~~~~~~~~~~~~~~~~~~ f\ll 1 (\D\ll B + \ehh)\cdot
\exp{- 2\pi \D\ll B}}}
\een{vanishingprim}
By definition of $\D\ll +$, the function $I\ll{31}(\D + \ehh)$ is greater than $K\ll{31}(\ehh)$ for $\D > \D\ll +$.
So if it were the case that $\D\ll 1 > \D\ll +$, then we would have the inequalities
\bbb
\D\ll n \geq \D\ll 1 > \D\ll +, ~~{\rm all}~n \geq 1\ ,
\xxn
I\ll{31} (\D\ll n + \ehh ) > K\ll{31} (\ehh), ~~{\rm all}~n \geq 1\ .
\een{inequalitiesprim}
We have one additional lemma to establish: that $\D\ll +$ is necessarily greater than $ {1\over{4\pi}} - \ehh$ if
$c$ and $\tilde{c}$ are greater than $1$.  It is straightforward to
check this property of $\D\ll +$ numerically, and we establish it
with an analytic proof in one subsection of the Appendix.
This property of $\D\ll +$ means that
\bbb
f\ll 1 (\D\ll n + \ehh ) = 2\pi \D\ll n + 2\pi\ehh  - \hh  > 0, ~~{\rm all}~n \geq 1\ .
\een{f1ineq}
From the inequalities \rr{inequalitiesprim} and \rr{f1ineq}, we could
infer that every term in the numerator and denominator of the
right hand side of \rr{vanishingprim} would have to be positive, so the equation
\rr{vanishingprim} would
be inconsistent.  Therefore our hypothesis $\D\ll 1 > \D\ll +$ cannot be true, and we conclude that
\bbb
\D\ll 1 \leq \D\ll +\ , Q.E.D.
\een{mainresult}

\subsection{Extended chiral algebras, and $c\leq1$}

We will comment briefly on the special cases we have excluded from
our considerations.

\heading{CFT with extended chiral algebras}

Our proof assumes that there does not exist an extended chiral algebra in the CFT -- that
is, that there does not exist an operator other than the identity with $h = 0$ or
$\tilde{h} = 0$.  Relaxing our assumptions to include such operators may lead to further interesting results.  In particular, the case $h = 0, \tilde{h} = 1$ (or vice versa) is interesting to consider: this is the case in which the CFT carries continuous global
current algebra symmetries.  In this case there are necessarily a tower of Virasoro primaries which have low dimension by virtue of being current algebra descendants of the identity.  (In the bulk interpretation, these states correspond to a gas of chiral gauge field
excitations confined to the boundary).  The interesting calculation in this case would be to derive an upper bound on the weight of the lowest nontrivial
operator that is primary with respect to the \it full \rm chiral algebra -- the Virasoro and current algebra pieces simultaneously.
When the current algebra
group is abelian, such a derivation could be interpreted as an upper bound on the mass of the lightest
charged state
in a theory of gravity, and could amount to a rigorous proof of the "weak gravity conjecture" \cite{weakgravity} of Arkani-Hamed \it et al. \rm for the case of negative cosmological constant.

Note also that the holomorphically factorized CFT of \cite{threedqgadscftwitten, hohnone, hohntwo} also contain very large chiral algebras, and the methods of our proof are not directly relevant to these.  In general, though, CFT with chiral algebras tend to have low-dimension operators (for instance, the chiral algebra itself), and it is likely that the best possible bound for $\D\ll 1$ will tend to be lower in a theory with a nontrivial chiral algebra than in a theory without, of the same central charge.  Note, for instance, that at large central charge, the upper bound in \cite{threedqgadscftwitten, hohnone, hohntwo} on the weight of the lowest Virasoro primary is lower than the bound derived in this paper, by a factor of two.  \edA{This is an interesting direction for further investigation.}

\heading{The special case of $c \leq 1$}

Our removal of \subA{Virasoro}{gravitational} descendants in the case $c > 1$
relied on the structure theorem for unitary representations of the
Virasoro algebra.  For $c \leq 1$ there are unitary Virasoro modules with
nontrivial structure that one must take into account if one wants to
extend the proof.

Conformal field theories with  $\ccr \leq 1$ but $\ccl > 1$ are not
classified, and may come in an infinite variety.  However
we would expect that there
should always be a large left-moving chiral algebra
in such theories, and the special issues involving extended chiral
algebras would become relevant.

The gravitational interpretation of theories with $0 < \ccr \leq 1$ and $\ccl$
very large is still open.  If we would like to describe the bulk
theory with a local Lagrangian, we would seem to
need to add a gravitational
Chern-Simons term \cite{tmg1, tmg2}, of the kind used in \cite{chiralgravity}.
The paper \cite{chiralgravity} only deals with the case in which one of
the two central charges strictly vanishes.  It is not clear how to generalize
\cite{chiralgravity} to the case where the left- and right- moving central
charges are unequal but both nonzero.  Other AdS/CFT dualities exist
\cite{murthy} in which
the left- and right-moving central charges are nonzero and unequal, and
described in the bulk by a perturbative string theory rather than a
local action.  However in all such examples the left- and right-moving
central charges are both large; neither is between 0 and 1.

In the case where both $\ccr$ and $\ccl$ are less than or equal to 1,
our understanding of the CFT is complete:
compact, unitary CFT
with $c < 1$ are completely classified\footnote{For a review of the classification of modular-invariant theories with $\ccr,\ccl < 1$, see \cite{dms}.} and it is possible to
inspect the operator spectra of these theories directly rather than
deriving a bound by abstract methods.
The range $0 < \ctot \leq 2$ represents
$AdS_3$ spaces with Planck-scale curvatures, so these cases are
exotic at best as theories of gravity in three dimensions.

\section{The gravitational interpretation of the upper bound on $\D\ll 1$}

In this section we turn to the gravitational interpretation of our CFT results.
We have derived an inequality that is completely universal in the set of two-dimensional conformal field theories, with some mild conditions:
unitarity, discreteness of the spectrum, and the condition that $\ccr ,\ccl > 1$, as
well as the absence of purely
left- or right-moving operators other than the stress tensor.

Such a universal inequality may have many interesting applications.
In particular, our bound is relevant for
the physics of gravity with negative cosmological constant.
The virtue of our approach is that we can derive a nontrivial, non-asymptotic constraint on the spectrum of a theory of gravity with negative cosmological constant.  In order to express our
bound in this form, we need only express the cosmological constant of the theory in terms of
the central charge of the corresponding CFT, and the mass of a state in
terms of the dimension of the corresponding local operator.

\subsection{Central charge and AdS radius}

In the case of the AdS$_3$/CFT$_2$ correspondence, the matching between the central charge of the CFT and the (negative) cosmological constant predates the undestanding of the AdS/CFT correspondence as
a dynamical principle.  Brown and Henneaux \cite{brownhenneaux} were able to identify the central charge in the Virasoro algebra of AdS$_3$ symmetries, based purely on the structure of the classical Poisson bracket algebra, leading to the identification
\bbb
\ccr  + \ccl = {3\over{ G\ll{\rm N} \sqrt{|\FIXITT|}}}\ .
\een{centralchargenewton}
It was later verified \cite{stromingercardy} that this is indeed the correct central charge for
the CFT that corresponds to the theory in the sense of \cite{maldacena}.

\subsection{Dimensions, masses, and rest energies}

We also wish to match the spectrum of massive objects with the spectrum of primary operators.  A primary state should be thought of as corresponding to a state at
rest with respect to the global time coordinate of AdS, because its energy cannot be lowered by acting with boost generators.  As always, we simplify the situation by assuming the absence of holomorphic primary operators.  (These would have a little group different from that of a massive particle in the bulk of AdS; therefore for small $\FIXITT = - L\uu{-2}$ they can only correspond to massless
states, which do not have a rest frame, or else to states which do not
propagate into the bulk of AdS at all.)

So we have the correspondence
\bbb
E\upp{rest} = {{\D}\over{L}}\ ,
\een{restenergy}
where $E\upp{rest}$ is the rest energy of an object in the bulk of AdS, and
$\D$ is the dimension of the primary operator.

For a minimally coupled massive scalar field of mass $m\ll s$, we
could use the dictionary of \cite{wittenadscft}, \cite{maldacenastrominger}:
\bbb
\D = 1 + \sqrt{1 + m\sqd\ll s L\sqd}
\xxn
\Leftrightarrow
\xxn
m\ll s = {1\over{L}}\cdot \sqrt{\D\sqd - 2 \D}\ .
\een{adscftdictionarydimensions}

The formula \rr{adscftdictionarydimensions} gives
\bbb
m\ll s\simeq {{\D  }\over L} + o\lrdd \D\uu 0 \rrdd
\eee
when $\D$ is large.  In the limit where $L\to\infty$ with $m\ll s$ held fixed, the order ${1\over{\D}}$ terms in the difference between $m\ll s$ and the rest energy of a massive excitation can be thought of as
the coupling of the massive field to the AdS curvature.

\subsection{Primaries and descendants}

As for the bulk interpretation of descendants of the primary with dimension $\D$, we follow \cite{threedqgadscftwitten} in interpreting these
as the original massive state in the bulk with boundary metric excitations added.

To be more precise, the states obtained by acting with $L\ll{-n}, \tilde{L}\ll{-n}$ with $n\geq 2$ correspond with creation operators for
quadrupole and higher modes of the metric; these are localized at spatial infinity and can be thought of loosely as "boundary gravitons" or "boundary metric excitations".  (No graviton states ever  propagate in the bulk in three dimensions.)  Acting with $L\ll{-1}$
and $\tilde{L}\ll{-1}$, on the other hand, can be thought of as exciting the dipole mode of the metric\edA{, which is pure gauge when applied to the
vacuum, but not pure gauge when applied to a state with a massive object in the bulk}.  In other words, the raising operators $L\ll{-1}$ and $\tilde{L}\ll{-1}$ boost the massive object in the bulk to a state of motion with higher energy.
So the primary states of the CFT correspond one to one with massive states in the bulk that are at rest, in which no boundary gravitons are excited.
Their descendants correspond to objects either in a nonzero state of motion, or with some
boundary gravitons excited, or both.

\subsection{Bulk interpretation of the upper bound on $\D\ll 1$}

We do not wish to assume that the lightest massive state is a scalar, nor
that it is necessarily described by a minimally coupled local field,
so we will use the formula \rr{restenergy} to interpret our bound \rr{mainresult} in
terms of the bulk physics in the flat-space limit.  Using formul\ae
~
\rr{centralchargenewton} and \rr{restenergy}, we can interpret \rr{mainresult} as saying that every consistent theory of quantum gravity with negative
cosmological constant $\FIXITT = - L\uu{-2}$ must necessarily have a massive
state in the bulk (with no boundary gravitons excited), with center-of-mass energy equal to $M\ll 1$, where
\bbb
M\ll 1 < M\ll +
\xxn
M\ll + \equiv
\lno\hilo {1\over L}~\D\ll + \rba\ll{\ctot = {{3 L}\over{G\ll N}}}\ .
\een{gravbound}
Of course we are assuming, as always, that the AdS radius $L$ is not so
small that $\ccr, \ccl \leq 1$.

We can now use our best linear upper bound on $\D\ll +$, as derived in the
Appendix and stated in \rr{bestlin}.  There, we show that
\bbb
\D\ll + < {{\ctot}\over{12}} + \d\ll 0\ ,
\een{bestlinearboundrecap}
where $\d\ll 0 \equiv 0.473695$.  In gravitational terms, this means that
\bbb
M\ll 1 \leq M\ll + < {1\over{4 G\ll N}} + {{\d\ll 0}\over L}\ .
\een{bestlingravboundwithcor}
In the flat-space limit $\FIXITT \to 0$, this says that
\bbb
M\ll 1 \leq {1\over{4 G\ll N}}\ .
\een{bestlingravbound}

This value of the mass $M\ll 1$ is suggestive.
The rest energy of the lightest BTZ black hole is ${1\over{8\, G\ll N}}$ above the
energy of the vacuum.\footnote{We should think of ${1\over{8\, G\ll N}}$ as the rest mass of the lightest
object in the spectrum of classical black holes, despite the fact that the lightest BTZ black
hole is often referred to as the "zero mass" BTZ black hole.}
The maximum possible value of $M\ll 1$ is twice that amount, so intuitively we may
say that, since the BTZ black hole exists as a state in every theory of 3D gravity and matter, then
there should always be a massive state at about that energy scale, even when quantum corrections are
taken into account.

We cannot, however, find any independent
bulk reasoning that could predict the coefficient of ${1\over{G\ll N}}$ in such an upper bound.
The tree-level mass of a Planck-scale black hole
need not have
any particular significance at the quantum level: small black holes would be expected
to receve $o(1)$ multiplicative renormalizations to their masses from virtual matter
particles.  It is not apparent how one could use bulk reasoning
to prove any upper bound on the quantum mass renormalization of the lightest
black hole, for a general theory of gravity coupled to matter.

The mass ${1\over{4\, G\ll N}}$ has been argued \cite{ashtekar} to have a special significance
when $\FIXITT = 0$, as the maximum value of the total energy of a collection of matter
coupled to gravity in 2+1 dimensions.
\footnote{We thank Lee Smolin for making us aware of this paper.}  The reason is
simple to understand: viewed from a long enough distance away, any collection of
matter with energy $M$ looks like a point particle, which creates a conical defecit in the
metric of $\Delta \phi = 16\pi G\ll N  M$.  For $M = {1\over{4\, G\ll N}}$
this means the space closes off entirely into a sphere, and for $M > {1\over{4\, G\ll N}}$
there is no consistent geometry at all.

For small but negative $\FIXITT$ the closing-off of the space can be avoided if
the collection of matter diffuse enough that the negative vacuum energy in any
region cancels or
overcancels the
positive energy carried by the matter.  This cancellation
cannot be achieved if the matter is made of
point particles of mass $M = {1\over{4\, G\ll N}}$ but it could
be achieved if the objects of $M = {1\over{4\, G\ll N}}$ were
composite objects such as solitons or strings.

We are not entirely
certain how to relate the result of \cite{ashtekar}
to the bound \rr{bestlingravboundwithcor}
the limit $\FIXITT\to 0$.
It may be that any theory of gravity and matter saturating \rr{bestlingravboundwithcor}
is necessarily very degenerate in the flat-space limit,
with dynamics that break up into a product of disjoint
systems with a small number of states in each one.  Alternately, the
result of \cite{ashtekar} may be a
sign that a better upper bound on $\D\ll 1$ can be proven
that is lower than $\D\ll +$ by
some numerical factor, in the limit $\ctot\to\infty$.
Such a conclusion would fit well with
the results \cite{hohnone, hohntwo, threedqgadscftwitten, mooreetal}.

\section{Conclusions}

In this note we have derived a rigorous upper bound on the scaling dimension
of the lowest primary operator (other than the identity) in a two-dimensional
conformal field theory \edA{that is invariant under the modular $S$-transformation}.  This bound is universal among all unitary conformal
field theories satisfying some very mild conditions:
$\ccr,\ccl > 1$ and the absence of purely left- or right-moving
operators beyond the components of the stress tensor itself.

%%%%%%%%%%%%%%%%%%%%%%%%%%%%%%%%%%%%%%%%%%%%%
%%%%%%%%%%%%%%%%%%%%%%%%%%%%%%%%%%%%%%%%%%%%%
\begin{figure}[htb]
\begin{center}
\includegraphics[width=5.0in,height=5.0in,angle=0]{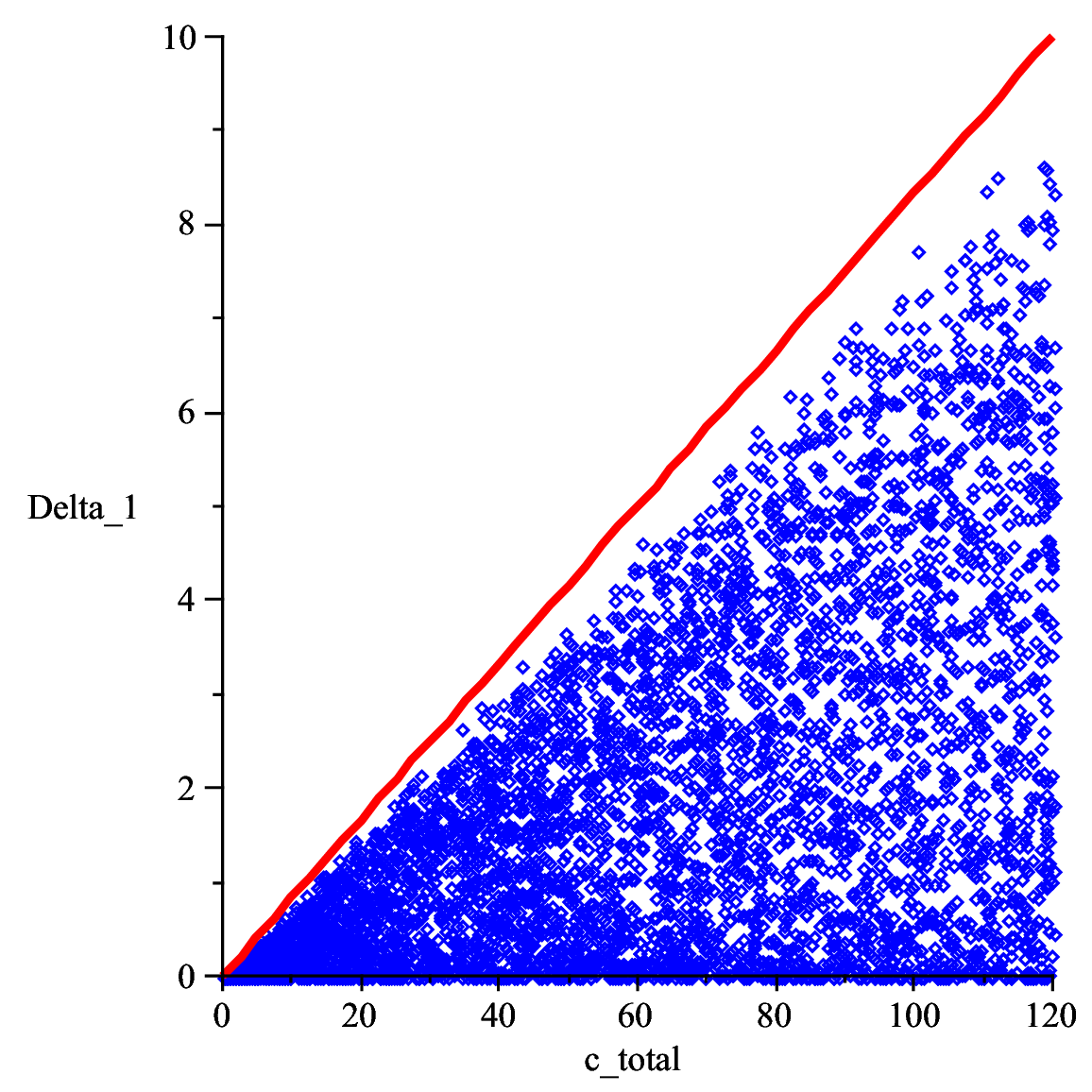}
\caption{In this paper we have proven that the distribution of unitary conformal
field theories in two dimensions looks something like the scatter plot above, where $\Delta_1$ is the weight of the lowest primary operator.  It is an
open question whether there exist CFT that saturate the bound at leading
order in $\ctot$, or whether further considerations could reduce the slope
of the red bounding line from ${1\over{12}}$, perhaps to as low as ${1\over{24}}$.}
\label{actualbound}
\end{center}
\end{figure}
%%%%%%%%%%%%%%%%%%%%%%%%%%%%%%%%%%%%%%%%%%%%%
%%%%%%%%%%%%%%%%%%%%%%%%%%%%%%%%%%%%%%%%%%%%%

As a warm-up, we also derived an upper bound for the scaling dimension of the lowest operator of any kind -- primary or descendant -- other than the identity.  For sufficiently large values of the total central charge, this version of the bound provides no information, since every theory contains a stress tensor, which has scaling dimension 2.  But for
$\ccl + \ccr$ less than $24 - {{18}\over{\pi}} \simeq 18.270$, even this rudimentary version of the bound does predict the existence of a primary of dimension less than 2.

One compelling open
question is to what extent $\D\ll + \sim {{\ctot}\over{12}}$
is the best bound possible at large central charge, in the full set of
2D conformal field theories satisfying our conditions.

On the one hand, there is
some reason to suspect that our bound could be improved.
In the holomorphically factorized case \cite{threedqgadscftwitten},
which includes the biggest gap in the spectrum of primaries
of any known class of examples, the
bound $\D\ll 1 \leq {{\ccr}\over{24}} + 1$
is a factor of two lower than the one we have derived, for
large $\ccr$.  The work of \cite{mooreetal} suggests the same bound in
another special class of theories with $(2,2)$ supersymmetry.  If these
special classes of CFT are typical of 2D CFT as a whole, then
it should be possible to cut our upper bound $\D\ll +$ in half.

We also observe that in the familiar cases of AdS$_3$/CFT$_2$ duality, there is
not only a single primary operator lying below $\D\ll +$, but many.
In fact, in every known example, there are an infinite number of primary
operators with $\D < \D\ll +$ in the limit $\ctot \to \infty$.  In terms
of the bulk, these operators
can be realized as strings, Kaluza-Klein modes of a decompactifying internal
dimension, or some other states of energy lower than ${1\over{4 G\ll N}}$.  This pattern
also suggests the possibility of improving on our inequality.

On the other hand, it may well be that the holomorphically factorized
case is a misleading guide to the upper bound on $\D\ll 1$
in the general case.  Holomorphically factorized CFT have many special
properties that are highly atypical of the set of CFT as a whole.
As an obvious example,
we note that factorized CFT always lack the chaos and thermalizing
behavior associated with true black holes.\footnote{In
a factorized CFT, amplitudes are automatically
periodic in time with period $2\pi$, in units where the
radius of the spatial circle is $1$.  Translated into
gravitational language, this means that every correlation function would
return precisely to itself after shifting the Lorentzian time coordinate of any one of the operator insertions
by the AdS time $L = |\FIXITT|\uu{-\hh}$.  Thermal correlators therefore cannot
decay exponentially as they would in the presence of black hole, for
a generic theory of gravity coupled to matter \cite{adsbh1, adsbh2, adsbh3}.}
The extreme specialness of holomorphically factorized
CFT, in this respect and others, suggests
that they may not ever realize the largest gap $\D\ll 1$ in primary operator dimensions that is achievable in general.

The other prominent examples \cite{mooreetal} in which there is
good evidence for a bound of $\D\ll 1\simeq {{\ctot}\over{24}}$, also have
atypical properties,
by virtue of their $(2,2)$-extended
supersymmetry.  The states of dimension $\D\ll 1 \simeq {{\ctot}\over{24}}$, that realize the proposed bound of \cite{mooreetal},
are BPS operators -- they are chiral primaries of the
(2,2) superalgebra.  Their dimensions, and the masses of the corresponding bulk states, are
protected by supersymmetry from any renormalizations.
For generic deformations of these theories
breaking all of their SUSY, we might expect that the mass of the state could receive
an $o(1)$ multiplicative
renormalization.  Such a renormalization could push the mass upwards by some factor from its a tree-level value of ${1\over{8\, G\ll N}}$,
possibly to as high as ${1\over{4\, G\ll N}}$.

Thus it is not clear at present whether
our bound $\D\ll 1 \leq \D\ll + \simeq {\ctot\over{12}}$
is the lowest achievable in the set of generic 2D CFT.
It would be interesting to know with confidence what the value best possible
bound is for large $\ctot$.  It may be possible to learn this optimal value, by
deriving an inequality for $\D\ll 1$ in the general case
together with a series of examples where $\D\ll 1$ saturates
the inequality in the limit where $\ctot$ is large.

The goal of our work was to understand how the rigorous holographic
definition of quantum gravity in terms of CFT
might generate universal predictions that would hold among \it all \rm
theories with an AdS ground state of a given size.
We have actually exceeded that goal: we have derived an upper bound on the mass of the lightest
massive state that is \it independent of the boundary conditions \rm in the limit where the
AdS becomes large.  That is, the universal upper bound on the lightest
mass approaches a
finite limit in Planck units, giving us a universal, falsifiable
prediction about local bulk physics, that does not
refer at all to the regulating AdS boundary condition.

The desirability of such robust
predictions has grown increasingly acute as our understanding of quantum gravity has developed:
just as we have come to understand that holography provides the "genetic code" of quantum gravity, we have simultaneously discovered an unimaginably vast and complex jungle of theories \cite{landscape} realizing that underlying code in myriad ways.
We are only now beginning to learn what family resemblances
the flora and fauna of this vast ecosystem have in common.

% HEER are addenda

\section{Acknowledgements}

The author would like to thank several people for valuable discussions, including Nima Arkani-Hamed, John Cardy,
Laurent Freidel,
Sergei Gukov, Petr Ho\v rava, Zohar Komargodski, Wei Li,  Michael Peskin,
Stephen Shenker, Lee Smolin
and Leonard Susskind.  We are particularly grateful
to Juan Maldacena for early conversations that contributed to the development of the ideas presented here, and
also for helpful
comments on the draft.  In addition, the author thanks the participants in the IPMU string theory group meetings
for raising the question of an upper bound on $\D\ll 1$ in a general 2D CFT, and the application
of such a bound to quantum gravity in AdS$_3$.  We are grateful to several institutions that
have provided us with hospitality during the completion of this work, including:
the Center for Cosmology and Particle Physics at NYU;
the Center for Theoretical Physics at MIT;
the Stanford ITP; the Perimeter Institute; the 2009 BIRS Workshop on Gauge Fields, Cosmology and Mathematical String Theory;
and the Center for Theoretical Physics at UC Berkeley.
This research was supported by the World Premier International Research Center Initiative,
MEXT, Japan; and by a Grant-in-Aid for Scientific Research (22740153) from the
Japan Society for Promotion of Science (JSPS).

%%%%%%%%%%%%%%%%%%%%%%%%%%%%%%%%%%%%%%%%%%%%%%%%%%%%%%%%%
%%%%%%%%%%%%%%%%%%%%%%%%%%%%%%%%%%%%%%%%%%%%%%%%%%%%%%%%%
%\renewcommand{\thefigure}{A-\arabic{equation}}
\setcounter{equation}{0}
%%\numberwithin{equation}{section}
%\appendix
%%%%%%%%%%%%%%%%%%%%%%%%%%%%%%%%%%%%%%%%%%%%%%%%%%%%%%%%%
%%%%%%%%%%%%%%%%%%%%%%%%%%%%%%%%%%%%%%%%%%%%%%%%%%%%%%%%%

%\renewcommand{\theequation}{A.\arabic{equation}}
% redefine the command that creates the equation no.
%\setcounter{equation}{0}  % reset counter

%%%%%%%%%%%%%%%%%%%%%%%%%%%%%%%%%%%%%%%%%%%%%%%%%%%%%%%%%%%%%%%%%%%%%%%%%%%%%
%
%  APPENDIX -- BEGINS HERE !!
%
%%%%%%%%%%%%%%%%%%%%%%%%%%%%%%%%%%%%%%%%%%%%%%%%%%%%%%%%%%%%%%%%%%%%%%%%%%%%%%%

%%%%%%%%%%%%%%%%%%%%%%%%%%%%%%%%%%%%%%%%%%%%%%%%%%%%%%%%%%%%%%%%%%%%%%%%%%%
\appendix{Properties of the function $\D\ll +(\ehh)$}

In this Appendix we will establish some facts about the function $\D\ll +(\ehh)$.  We only deal with CFT for which $\ccr, \ccl >1$,
so we will always restrict the domain of definition of $\D\ll +(\ehh)$ to
$\ehh \leq 0$, or equivalently $\ctot > 2$.

\subsection{Definition of $\D\ll +$}

Recall that we have defined $\D\ll +(\ehh)$ to be the largest real root
$\D$ of the cubic polynomial
\bbb
P_{31}(\Delta) \equiv
f\ll 3( \D + \ehh) - K\ll {31}(\ehh)
f\ll 1(\D + \ehh) = {1\over{b\ll 1(\ehh)}} ~F\ll{31}(\Delta)\ ,
\xxx
F\ll{31}(\Delta)\equiv
 b\ll 1(\ehh) f\ll 3(\D + \ehh) -  b\ll 3(\ehh) f\ll 1(\D + \ehh)\ .
\eee
We have defined $K\ll{31}$ in \rr{kdef}, in the main body of the paper.

The explicit, analytic expressions for the polynomials $f\ll{1,3}$ and $b\ll{1,3}$ are:
\bbb
f\ll 1 (z) = (2\pi z) - \hh\ ,
\xxx
f\ll 3 (z) = (2\pi z)\uu 3 - {9\over 2}\, (2\pi z)\sqd + \lrdd {{41}\over 8} + 6\, r\ll{20}
\rrdd (2\pi z) - \lrdd {{17}\over{16}} + 3\, r\ll{20} \rrdd\ ,
\xxx
b\ll 1 (z) = f\ll 1(z) - 2\, \exp{-2\pi}\, f\ll 1(z + 1) + \exp{-4\pi}\, f\ll 1(z + 2)\ ,
\xxx
b\ll 3 (z) = f\ll 3(z) - 2\, \exp{-2\pi}\, f\ll 3(z + 1) + \exp{-4\pi}\, f\ll 3(z + 2)\ .
\eee
It is a straighforward exercise to establish that the constant $r\ll{20}$
defined in \rr{rtwozerodef}
does not appear in the expression $b\ll 1(x) f\ll 3(y) - b\ll 3(x) f\ll 1(y)$
for general $x,y$.  The expression $F\ll{31}(\Delta)$ is of this general form,
so we will never need to use the value of $r\ll{20}$.  Thought of as
a polynomial in two variables $(\D,\ehh)$, the
coefficients in expression $F\ll{31}$ involve only $\pi$ and $\exp{-2\pi}$.

For completeness, we wil write the full expression for $F\ll{31}$, for
general $\ehh$.
\def\leftblank{}
\def\rightblank{}
\bbb
F\ll{31}(\Delta) =  \sum\ll{m = 0}\uu 3 \sum\ll{n = 0}\uu
{4 - m} A\ll{mn} \ehh\uu n \Delta\uu m
\eee
with the coefficients $A\ll{mn}$ as follows:
\bbb
A\ll{31} = \leftblank  16 \pi ^4 ~ \left ( 1-\exp{- 2\pi} \right )^2
\rightblank
\xxx
A\ll{30} =
- \leftblank  4 \pi ^3  \left ( 1-\exp{- 2\pi} \right )
  \left [\hilo  1+(8 \pi  - 1)
 \exp{- 2\pi} ~\right ] \rightblank
\xxx
A\ll {22} =
\leftblank  48 \pi ^4 (1-\exp{- 2\pi})^2\rightblank
\xxx
A\ll{21} =
-\leftblank  48 \pi ^3
 \left ( 1-\exp{- 2\pi} \right )
 \left [ \hilo
1+(2 \pi - 1) \exp{- 2\pi} ~ \right ] \rightblank
\xxx
A\ll{20} =
\leftblank  9 \pi ^2  \left (
1-\exp{- 2\pi} \right ) \left [\hilo 1+ \left ( 8 \pi - 1 \right )
 \exp{- 2\pi}~\right ] \rightblank
\xxx
A\ll{13} = \leftblank  32 \pi ^4  \left ( 1-\exp{- 2\pi}
\right )^2\rightblank
\xxx
A\ll{12} = \leftblank  -48 \pi ^3  \left ( 1-\exp{- 2\pi} \right )^2
\rightblank
\xxx
A\ll{11} = \leftblank  6 \pi ^2 \left [ \hilo
3+2 \left(8 \pi ^2 - 3\right) \exp{- 2\pi}-\left(32 \pi ^2 - 3 \right)
\exp{- 4\pi}
\right ]
\rightblank
\xxx
A\ll{10} = - \leftblank  \pi  \left [ \hilo 3-\left(32 \pi ^3 - 72 \pi ^2
+ 6\right) \exp{- 2\pi}+\left(128 \pi ^3  -144 \pi ^2+
 3 \right) \exp{- 4\pi}
\right ]
\rightblank
\xxx
A\ll{04} = 0
\xxx
A\ll{03} = \leftblank  64 \pi ^4 ~ \exp{- 2\pi}~(1-\exp{- 2\pi}) \rightblank
\xxx
A\ll{02} = \leftblank  96 \pi ^3 ~ \exp{- 2\pi}
\lsqq
(\pi -1) -(2\pi - 1) ~ \exp{- 2\pi} \rsqq \rightblank
\xxx
A\ll{01} = \leftblank  4 \pi ^2 ~ \exp{- 2\pi}
\left [ \hilo (8\pi\sqd - 24\pi + 9)
- ( 32 \pi\sqd - 48\pi +9) ~\exp{- 2\pi}
\right ]
\rightblank
\xxx
A\ll{00} = - \leftblank  2 \pi  ~ \exp{- 2\pi} ~
\left [ \hilo (4\pi\sqd -9 \pi + 3) - (16\pi\sqd - 18\pi + 3)~\exp{- 2\pi}
~ \right ]
\rightblank
\eee

\subsection{$\D\ll +$ is a smooth function of $\ctot$}

We will now show that $\D\ll +$ is continuous as a function of $\ctot$ in the range of
interest, $\ctot\in [2,\infty)$.  $\D\ll +$ is defined implicitly as the largest root $\Delta$ of
the polynomial $P\ll{31}(\Delta, \ehh)$, with $\ehh \equiv {{2 - \ctot}\over{24}}$.
The value of a root of a polynomial depends smoothly on its coefficients, except when two
roots become coincident.  (This assumes the leading coefficient of the polynomial is
constant, as in this case.)
So in order to show that $\D\ll +$ depends smoothly on $\ctot$, we need to show that the
coefficients in $P\ll{31}$ are smooth functions of $\ctot$, and that the roots of $P\ll{31}$ are all
distinct.

The coefficients of $P\ll{31}$ depend smoothly on $\ehh$ as long as the rational function
$K\ll{31}(\ehh) = {{b\ll 3(\ehh)}\over{b\ll 1(\ehh)}}$ does.  The denominator $b\ll 1(\ehh)$, is
given explicitly by
\bbb
b\ll 1 (\ehh) = 2\pi(1 - \exp{- 2\pi})\sqd~\ehh - \hh (1 - \exp{- 2\pi})\sqd - 4\pi \exp{- 2\pi}
(1 - \exp{- 2\pi})\ .
\eee
The second and third terms are negative, and the first is negative for $\ctot > 2$ and vanishes
for $\ctot =2$.  So $b\ll 1(\ehh)$ can
never vanish, and the coefficients of $P\ll{31}$ depend smoothly on $\ctot$ for $\ctot \in
[2, \infty]$.

So the only possible way $\D\ll +$ could be non-smooth would be if $P\ll {31}$ had
coincident roots for some value of $\ctot$.
Fortunately $P\ll{31}$ is only
cubic as a function of $\D$, so it is straightforward to check for coincident roots.
Let us put $P\ll{31}$ into canonical form for a cubic polynomial by changing variables
so that the quadratic term vanishes and the coefficient of the leading term is 1.

The two leading terms of $P\ll{31}$ are
\bbb
P\ll{31}(\Delta, \ehh) = 8\pi\uu 3 \lsqq \D\uu 3 + \lrdd 3 \ehh - {9\over{4\pi}} \rrdd
\D\sqd + o\lrdd \D \rrdd \rsqq
\eee
so let us take
\bbb
x \equiv 2\pi(\D +  \ehh) - {3\over 2 } \llsk\llsk \D = {x\over{2\pi}} - \ehh + {3\over{4\pi}} \ .
\eee
Then
\bbb
P\ll{31} (\Delta,\ehh) = f\ll 3 ({{x + {3\over 2}}\over{2\pi}})
- K\ll{31} (\ehh) f\ll 1({{x + {3\over 2}}\over{2\pi}})
\xxx
= x\uu 3 + C\ll 1 (\ehh) x + C\ll 0 (\ehh)
\eee
with
\bbb
C\ll 1(\ehh) \equiv - K\ll{31}(\ehh) + 6 r\ll{20} - {{13}\over 8}
\xxx
C\ll 0(\ehh) \equiv - K\ll{31}(\ehh) + 6 r\ll{20} - {1\over 8} = C\ll 1(\ehh) + {3\over 2}
\eee
Note that the constant $r\ll {20}$ drops out of
the combinations $C\ll 1(\ehh)$ and $C\ll 0(\ehh)$.

Now we make use of the well-known formula for the discriminant of a cubic polynomial
with leading term equal to 1 and vanishing quadratic term.  The discriminant is
given by
\bbb
{\rm Disc}(\ehh) \equiv 4 C\ll 1 \uu 3
(\ehh)+ 27 C\ll 0\sqd(\ehh)
\eee
As a function of $x$, the roots of $P\ll{31}(\Delta, \ehh)$ are distinct when
${\rm Disc}(\ehh)$ vanishes.  Using $C\ll 1 = C\ll 0 - {3\over 2}$, we can write
\bbb
{\rm Disc}(\ehh) = {\bf D}(C\ll 0(\ehh))\ ,
\eee
where
\bbb
{\bf D}(y) \equiv 4 y\uu 3 + 9 y\sqd + 27 y - {{27}\over 2}
\eee
Using Cardano's formula, we can find all the roots of ${\bf D}(y)$.  Two of the roots
are complex, and the real root is
\bbb
y\st \equiv {3\over 4} \lsqq -1 + \lrdd 6 \sqrt{3} + 9 \rrdd\uu{1\over 3}
- \lrdd 6 \sqrt{3} - 9 \rrdd\uu{1\over 3} \rsqq \simeq 0.427505 + o \lrdd 10\uu{-7} \rrdd\ .
\eee
So the polynomial $P\ll{31}(\Delta, \ehh)$ can have a double zero only if
\bbb
C\ll 0 (\ehh) = - K\ll{31}(\ehh) + 6 r\ll{20} - {1\over 8} = y\st
\xxx
\Leftrightarrow
\xxx
\lsqq b\ll 3(\ehh) - 6 r\ll{20} b\ll 1(\ehh) \rsqq
 + (y\st + {1\over 8})  b\ll 1(\ehh) = 0
\eee
This equation has three real roots, at
\bbb
\hat{E}\ll 0\uu{*(1)} =  0.0821971 + o \lrdd 10\uu{-8} \rrdd\ ,
\xxx
\hat{E}\ll 0\uu{*(2)} = 0.184241 + o\lrdd 10\uu{-7} \rrdd\ ,
\xxx
\hat{E}\ll 0\uu{*(3)} = 0.460984 + o \lrdd 10\uu{-7} \rrdd\ .
\eee
In particular, all roots $\hat{E}\ll 0\st$ are positive.  But we are only interested in
$\ctot  > 2$, in which case $\ehh$ is negative or zero.  It follows that the
polynomial $P\ll{31}(\D,\ehh)$ has no multiple roots (as a function of $\D$) for
$\ctot \in[2,\infty)$.  We conclude that in this range
the function $\D\ll +(\ehh)$ depends smoothly on $\ctot$, \it Q. E. D. \rm

\subsection{The function $\D\ll +$ is greater than ${1\over{4\pi}} - \ehh$ for
$\ctot > 2$}

Now we wish to prove a lemma of which we make use in the body of the paper.  We would
like to show that $\D\ll +(\ehh) + \ehh > {1\over{4\pi}}$ for all $\ctot >  2$.
In order to show this,
it will be convenient to work in terms of the variable $x\equiv 2\pi(\D + \ehh) - {3\over 2}$,
in terms of which the polynomial $P\ll{31}(\D,\ehh)$ takes the canonical form
\bbb
P\ll{31}({{x + {3\over 2}}\over{2\pi}} - \ehh,\ehh) = x\uu 3 + C\ll 1(\ehh) x + C\ll 0(\ehh) \,
\een{definingpoly}

Define $x\ll +$ to be the largest root of this polynomial.
In terms of $x\ll +$, we need to establish that $x\ll + \geq -1$ at any
local minimum of $\D\ll +$.  The function $x\ll +(\ehh)$ could
develop a critical point in one of two ways: either the coefficient
functions $C\ll 0(\ehh)$ and $C\ll 1(\ehh)$ could have a critical point as a function
of $\ehh$, or the largest root $x\ll +$ polynomial $x\uu 3 + C\ll 1 x + C\ll 0$ could have a critical point
as a function of $C\ll 0$.  We will show that
the former possibility cannot occur for
$\ctot > 2$, and the latter possibility can never occur at all.

The condition for the functions $C\ll{0,1}(\ehh)$ to have a critical point is
\bbb
K\pr\ll{31}(\ehh) = 0 \llsk\Leftrightarrow b\pr\ll 3(\ehh) b\ll 1 (\ehh) - b\ll 1\pr(\ehh) b\ll 3
(\ehh)
\xxx
\simeq 3093.87 \ehh\uu 3 - 1511.95 \ehh\sqd + 187.515 \ehh - 6.9832\ .
\eee
The coefficients of this polynomial are strictly alternating, so it can have no negative roots.
It follows that $C\ll 0(\ehh), C\ll 1(\ehh)$ never have critical points as a function of
$\ctot$ for $\ctot > 2$.

Next we will show that $x\ll +$ can never have a critical point as
a function of $C\ll 0$.  Let $x\ll{1,2,3}$ be the three solutions to the equation
\bbb
= x\uu 3 + C\ll 1 x + C\ll 0
\xxx
= x\uu 3 + \lrdd C\ll 0 - {3\over 2} \rrdd x + C\ll 0\ .
\een{rewrittencubic}
These roots can be thought of as implicit functions of $C\ll 0$, and they satisfy
\bbb
x\ll 1 + x\ll 2 + x\ll 3 = 0
\xxx
x\ll 1 x\ll 2 + x\ll 1 x\ll 3 + x\ll 2 x\ll 3 = C\ll 0 - {3\over 2}
\xxx
x\ll 1 x\ll 2 x\ll 3 = - C\ll 0 \ .
\eee
Now suppose there is some value $C\ll 0\st$ of $C\ll 0$ such that one of the roots, say $x\ll 3$,
has a critical point as a function of $C\ll 0$.  Then at $C\ll 0 = C\ll 0\st$, we have
\bbb
\dot{x}\ll 3 = 0
\xxx
\dot{x}\ll 1 + \dot{x}\ll 2 = 0
\xxx
x\ll 1 \dot{x}\ll 2 + x\ll 2 \dot{x}\ll 1 = 1
\xxx
(x\ll 1 \dot{x}\ll 2 + x\ll 2 \dot{x}\ll 1) x\ll 3 = -1\ .
\eee
We conclude that any root $ x\ll 3$ of \rr{definingpoly}
that is a local extremum as a function
of $C\ll 0$ must necessarily take the value $x\ll 3 = -1$.
But if we plug this value back into the defining equation, we find
\bbb
0 = x\ll 3 \uu 3 + C\ll 1 x\ll 3 + C\ll 0 = (C\ll 0 - C\ll 1) - 1 = {3\over 2} - 1 = \hh\ ,
\eee
leading to a contradiction.  So no root of the defining polynomial can ever be equal to $-1$, and therefore
$x\ll +$ is monotonic as a function of $C\ll 0$.  Taken together with the result that
$C\ll 0$ is monotonic as a function of $\ctot$ for $\ctot > 2$, this implies
that $x\ll +$ is monotonic as a function of $\ctot$ in the same range.  We conclude that
$\D\ll +(\ehh) + \ehh$ is monotonic as a function of $\ctot$ for $\ctot > 2$.

We will see in the next subsection that $\D\ll +(\ehh)$ goes as $- 2 \ehh + o(\ctot\uu 0)
= {{\ctot}\over{12}} + o(\ctot\uu 0)$ at large $\ctot$, so $\D\ll +$ is monotonically
increasing as a function of $\ctot$ in the range of interest.

So we have learned that $\D\ll +(\ehh)$ is always greater than ${1\over{4\pi}}$ in the range
$\ctot > 2$.  It follows that $\D\ll +(\ehh) + \ehh$ is an increaing function of
$\ctot$ for $\ctot > 2$.  The value of $\D\ll +(\ehh) + \ehh$ at $\ctot = 2$ is
$\D\ll +(0) = 0.615286 + o \lrdd 10\uu{-7}\rrdd$, so we conclude
\bbb
\D\ll + (\ehh) + \ehh > \D\ll +(0)> {1\over{4\pi}}, ~Q.~E.~D.
\eee

\subsection{Behavior of $\D\ll +$ for large central charge}

Now let us take the total central charge $\ctot $ to
be large and positive.  Taking $\ctot \to + \infty$ means taking
$\ehh \equiv E\ll 0 + {1\over{12}} =
{{2 - \ctot}\over{24}}$ to $-\infty$.  In this limit
it is easy to see that $\D\ll +$ is proportional to $\ctot$, plus
corrections of order $\ctot\uu 0$.  To see this, it is useful to
expand $\D\ll +$ as a series at large central charge:
\bbb
\D\ll + \equiv \sum\ll{a = -1}\uu{\infty} \d\ll{-a} \lrdd {{\ctot}\over{24}}
\rrdd \uu{-a}\ .
\een{expansion}
The defining property of $\D\ll +$ is that it satisfies
\bbb
F\ll{31} (\D\ll +, \ehh)  =0\ ,
\eee
and that $\D\ll +$ is the largest real value with that property, for a
given value of $\ctot$.  Using the deinifiton of $\ehh$ in terms
of $\ctot$ and the expansion \rr{expansion} or $\D\ll +$,
we can expand $F\ll{31}(\D\ll +, \ehh)$ to arbitrary order in
${1\over{\ctot}}$,
and solve for the universal numerical coefficients $\d\ll{-a}$.
To leading order in $\ctot$, we thus obtain:
\bbb
F\ll{31} (\D\ll +,\ehh) = - {{\pi\uu 4 (1 - \exp{- 2\pi})\sqd}\over{20736}}
~\lrdd \hilo \d\ll 1\uu 3 - 3 \d\ll 1\sqd + 2\d\ll 1~\rrdd~\ctot\uu 4 + o\lrdd\ctot\uu 3
\rrdd \ .
\een{f31leadingorder}
We conclude that $\d\ll 1$ is the largest of the three roots of $\d\ll 1
\uu 3 - 3 \d\ll 1\sqd + 2\d\ll 1 = \d(\d\ll 1 - 1)(\d\ll 1 - 2)$.  So
$\d\ll 1 = +2$, which means
\bbb
\D\ll + = {\ctot\over{12}} + \d\ll 0 + o\lrdd \ctot\uu{-1} \rrdd\ .
\eee
To determine $\d\ll 0$ we expand $F\ll{31}$ to
order $\ctot\uu 3$.  Fixing $\d\ll 1 = +2$, we find
\bbb
F\ll{31} = - \lrdd\hilo {{\pi\uu 4 (1 - \exp{- 2\pi})\sqd}\over{432}} ~\rrdd
\ctot\uu 3~\lsqq\hilo \d\ll 0 - {{(12 - \pi) + (13\pi - 12)\exp{- 2\pi}}
\over{6\pi~(1 - \exp{-2\pi}) }}  ~\rsqq  + o\lrdd  \ctot\sqd \rrdd\ .
\eee
This determines the coefficient $\d\ll 0$ to be
\bbb
\d\ll 0 \equiv {{(12 - \pi) + (13\pi - 12)\exp{- 2\pi}}
\over{6\pi~(1 - \exp{-2\pi}) }} \simeq 0.473695 + o\lrdd \hilo 10\uu{-7}~\rrdd
\ .
\eee

We could easily determine the higher coefficients $\d\ll{-1},\d\ll{-2},\cdots$
to arbitrary order.  However we will not bother to derive any
coefficients beyond $\d\ll 0$, for a simple reason: There is no guarantee that
the bound we have derived is the best possible, even asymptotically
at large $\ctot$.  If the $\ctot\uu {-a}$ term in
$\D\ll +$ is not the lowest possible value that can be obtained by
any method, then the $\ctot\uu{-(a+1)}$ term
will not be relevant at all.  We are not even certain if our leading
expression $\D\ll + \simeq {{\ctot}\over{12}}$ is the best possible
upper bound at large $\ctot$, which means the finite correction $\d\ll 0$
may not be meaningful.  Even if it is, it seems quite unlikely that
${{\ctot}\over{12}} + \d\ll 0$ is the best possible
upper bound to order $\ctot\uu 0$.  Unless we have some reason to believe
that there exist actual CFT with total central charge $\ctot$
that can attain the values $\D\ll 1 = {{\ctot}\over{12}} + \d\ll 0
+ o(\ctot\uu{-1})$, there is nothing to be gained
in carrying the expansion of $\D\ll +$ to order $\ctot\uu{-1}$.

\subsection{The function $\D\ll +$ is bounded above by ${{\ctot}\over{12}}
+ 0.473695$.}

To extract the simplest possible conclusions from our inequality,
we would like to find a linear function $a~\ctot + k$
that is always greater than or
equal to $\D\ll +$ whenever $\ctot$ is greater than
2.  We would also like this function to be a good approximation to $\D\ll +$
at large $\ctot$, so we will take $a$ to be equal to $+{1\over{12}}$.

Let us find the lowest possible number
$k$ (independent of $\ctot$) such that $\D\ll +(\ehh) \leq {{\ctot}\over{12}}
 + k$ for all $\ctot > 2$.  It is convenient to rewrite this condition as
 $\D\ll + \leq - 2 \ehh + k\pr$, with $k\pr = k + {1\over 6}$, so that we can
 use the variable $\ehh$.

 If $k\pr$ is the smallest possible number such that the inequality is satisfied,
 then one of three possibilities holds:
 \begin{itemize}
 \item{The inequality could be saturated asymptotically as $\ehh \to - \infty$.}
 \item{The inequality could be saturated at $\ehh = 0$.}
 \item{The inequality could be saturated at some value of $\ehh$ in between $0$ and $-\infty$.}
 \end{itemize}
 Let us eliminate the third possibility.  Suppose there were some value $\ehh\upp{sat}$
 between $0$ and $-\infty$ such that the inequality is saturated.  Then the line
 $\D = - 2 \ehh + k\pr$ must be tangent to the curve $\D = \D\ll + (\ehh)$ at the point
 $(\ehh\upp{sat} , \D\ll + (\ehh\upp{sat}))$.  That is, the derivative $\D\pr\ll +(\ehh)$
 must be equal to $-2$ at the point $\ehh = \ehh\upp{sat}$.

 In terms of the variable $x\ll +\equiv 2\pi(\D\ll + + \ehh) - {3\over 2}$, we have
 the defining equation
 \bbb
 x\ll + \uu 3 + C\ll 1 x\ll +  + C\ll 0 = 0\ ,
 \een{stilldefining}
 and we wish to search for values of $\ehh$ such that $x\pr\ll +(\ehh) = - 2\pi$.
 The functions $x\ll +,C\ll 0, C\ll 1$ all vary with $\ehh$; using a prime to
 denote differentiation with respect to $\ehh$ we have
 \bbb
 3 x\ll +\sqd x\pr\ll + + C\ll 1\pr x\ll + + C\ll 1 x\ll + \pr + C\ll 0\pr = 0\ .
 \eee
 For the particular values $x\ll +\upp{sat}, \ehh\upp{sat}$ where the equation $x\pr = -2\pi$,
 so we have
 \bbb
\lno\hilo  - 6\pi x\ll +\sqd + C\pr\ll 1 x\ll +
 - 2\pi~C\ll 1  + C\ll 0\pr \rba\ll{\ehh = \ehh\upp{sat}} = 0\ .
 \een{firstvan}
 Multiplying by $x\ll +$ and using the defining equation \rr{stilldefining}, we obtain
 \bbb
 \lno\hilo   C\pr\ll 1 x\ll +\sqd + \lrdd 4\pi  C\ll 1  + C\ll 0 \pr \rrdd x\ll + +6\pi C\ll 0
  \rba\ll{\ehh = \ehh\upp{sat}} = 0\ .
 \eee
Combining the two to cancel the $x\ll +\sqd$ term, and solving for $x\ll +$, we get
\bbb
x\ll + \upp{sat} = \lno\hilo {{2\pi C\ll 1 C\pr\ll 1 - C\ll 0 \pr C\ll 1\pr - 36 \pi\sqd C\ll 0}\over
{C\ll 1\uu{\pr 2} + 6\pi C\ll 0\pr + 24\pi\sqd C\ll 1}} \rba\ll{\ehh = \ehh\upp{sat}}
\eee
Plugging back into \rr{firstvan}, we obtain
\bbb
- C\ll 0\pr{}\uu 3 + C\ll 0 C\ll 1\pr{}\uu 3 - 6 \pi C\ll 0\pr{}
\sqd C\ll 1 + 216 \pi\uu 3 C\ll 0{}\sqd
\xxx
+ C\ll 0\pr C\ll 1 \pr \lrdd - C\ll 1 C\ll 1\pr + 18 \pi C\ll 0\rrdd
+ 2\pi C\ll 1\sqd \lrdd C\ll 1\pr{}\sqd + 16\pi\sqd C\ll 1 \rrdd  = 0
\eee
Substituting the actual values of $C\ll 1(\ehh)$ and $C\ll 0(\ehh)$ into this equation, the polynomial above can be factorized as
\bbb
1.82307 \times 10\uu{19} \cdot (6716.62~\ehh - 559.624)\uu{-7}
\xxx
\cdot (-1.02204 + \ehh) (-0.538217 + \ehh) (-0.0833194 + \ehh)
\xxx
\cdot (0.00793973 - 0.172323 \ehh + \ehh^2 ) (0.00701421 - 0.167214 \ehh + \ehh^2 )
\xxx
\cdot (0.0069072 - 0.165724 \ehh + \ehh^2 ) (0.0164746 - 0.0886746 \ehh + \ehh^2 )
\xxx
\cdot \lsqq 3.32249 \times 10\uu {13}   + (\ehh + 0.0163377)\sqd  \rsqq
\eee
%%%%%%%%%%%%%%%%%%%%%%%%%%%%%%%
%
% I factorized that polynomial by HAND, using PEN AND PAPER -- I **DIDN'T** USE MATHEMATICA!
%
% And it only took me about 5 minutes!
%
% Damn, I'm smart!
%
%%%%%%%%%%%%%%%%%%%%%%%%%%%%%
For $\ctot > 2$, we have $\ehh \leq 0$ and every
term in the above expression is nonvanishing.
We conclude that the function $\D\ll + (\ehh) + 2 \ehh$ has no critical points for $c > 2$.

So the coefficient $k$ that optimizes the bound $\D\ll + \leq {{\ctot}\over{12}} + k$
in the range $\ctot \in[2,+\infty)$ must occur at one of
the endpoints -- either at $\ctot = 2$ or in the limit $\ctot \to + \infty$.

So that means
\bbb
k = {\rm max} \lrdd\hilo \D\ll +(0) - {1\over 6}~~,
~~ \lim\ll{\ctot \to \infty}
\D\ll +({{2 - \ctot}\over{24}}) - {{\ctot}\over{12}} \rrdd
\eee
From the previous section, we know that $\lim\ll{\ctot \to \infty}
\D\ll +({{2 - \ctot}\over{24}}) - {{\ctot}\over{12}} = \d\ll 0
= 0.473695$.  We can check that the value of $\D\ll +(\ehh)$ at $\ctot = 2$
is
\bbb
\D\ll +(0) = 0.615286
\eee
so
\bbb
\D\ll +(0) - {1\over 6} = 0.448619\ ,
\eee
which is less than $\d\ll 0$.  So indeed $k = \d\ll 0$, and our best
linear bound is
\bbb
\D\ll 1 \leq \D\ll +(\ehh) \leq {{\ctot}\over{12}} + \d\ll 0 = {{\ctot}\over{12}} + 0.473695\ .
\een{bestlin}
In other words, retaining only the first two terms in the
asymptotic expansion of $\D\ll +$ at large $\ctot$ gives a linear function
that bounds $\D\ll +$ above uniformly on the semi-infinite interval of
interest, $\ctot\in [2,\infty)$.

We also point out that the value of ${3\over{2\pi}}$ is $0.477465$, so our bound on
primaries is very slightly better than the warm-up bound we derived for
the lowest-dimension nontrivial operator in general.

%%%%%%%%%%%%%%%%%%%%%%%%%%%%%%%%%%%%%%%%%%%%%%%%%%%%%%%%%%%%%%%%%%%%%%%%%%%

%ENDD

\end{document}